\documentclass[aps,rmp,reprint,amsmath,amssymb,graphicx,longbibliography]{revtex4-1}
\usepackage[caption=false]{subfig}
\usepackage{amsmath,graphicx}
\usepackage{amsmath,amssymb,mathrsfs,esint}
\usepackage{multirow}
\usepackage{algorithm}
\usepackage[noend]{algpseudocode}
\usepackage{comment}
\usepackage{tabularx}
\usepackage{bm}
\usepackage[normalem]{ulem}
\usepackage[bookmarks=true,
   colorlinks=true,
   linkcolor=blue,
   urlcolor=blue,
   citecolor=blue,
   bookmarks=true,
   hyperindex=true
]{hyperref}
\usepackage{tikz, comment, adjustbox, multirow}
\usepackage{braket}
\usepackage{bbold}
\usetikzlibrary{positioning}

\usetikzlibrary{quantikz2}
\usepackage{nicematrix}

\newcommand{\nqb}{\ensuremath{b}}

\begin{document}

\title{Qudits for decomposing multiqubit gates and realizing quantum algorithms}

\author{Evgeniy O. Kiktenko}
\email{evgeniy.kiktenko@gmail.com}
\affiliation{Russian Quantum Center, Skolkovo, Moscow 121205, Russia}
\affiliation{National University of Science and Technology ``MISIS'', Moscow 119049, Russia}

\author{Anastasiia S. Nikolaeva}
\email{anastasiia.nikolaeva21@gmail.com}
\affiliation{Russian Quantum Center, Skolkovo, Moscow 121205, Russia}
\affiliation{National University of Science and Technology ``MISIS'', Moscow 119049, Russia}

\author{Aleksey K. Fedorov}
\email{lex1026@gmail.com}
\affiliation{Russian Quantum Center, Skolkovo, Moscow 121205, Russia}
\affiliation{National University of Science and Technology ``MISIS'', Moscow 119049, Russia}

\date{\today{}}

\begin{abstract}
The paradigm behind digital quantum computing inherits the idea of using binary information processing. Nature in fact gives much more rich structures of physical objects that can be used for encoding information, which is especially interesting in the quantum-mechanical domain. In this
Colloquium several ideas are reviewed that indicate how multilevel quantum systems, also known as qudits, can be used for efficient realization of quantum algorithms, which are represented via standard qubit circuits. The focus in the Colloquium is on techniques for leveraging qudits for simplifying decomposition of multiqubit gates and for compressing quantum information by encoding multiple qubits in a single qudit. As discussed in the Colloquium, these approaches can be efficiently combined. This allows a reduction in the number of entangling (two-body) operations
and the number of quantum information carriers used compared to straightforward qubit realizations. These theoretical schemes can be implemented with quantum computing platforms
of various natures, such as trapped ions, neutral atoms, superconducting junctions, quantum light,
spin systems, and molecules. 
The Colloquium concludes by summarizing a set of open problems whose resolution will be an important further step toward employing universal qudit-based processors for running qubit algorithms.
\end{abstract}

\maketitle

\tableofcontents

\section{Introduction}
\label{intro}

The idea behind quantum computing is closely related to the fundamental problem of understanding the limit capabilities of real physical devices in solving computational problems~\cite{Markov2014}. 
Observations by Feynman~\cite{Feynman1982,Feynman1986} that classically intractable problems in physics and chemistry can be solved with a computer based on quantum-mechanical principles, 
as well as a much more rich structure of information encoding in the quantum domain emphasized by Manin~\cite{Manin1980}, have become the cornerstone concepts of quantum computing. 
Further formalizations of these concepts in terms of the quantum Turing machine by \cite{Deutsch1985} led to the Church-Turing-Deutsch principle. 
Leveraging complex entangled states~\cite{Preskill2012}, which can be efficiently prepared (at least, theoretically) in real physical devices, 
is believed to be a crucial ingredient of quantum computing that ensures their potential speedup in solving certain classes of computational problems. 
Examples of such problems include integer factorization~\cite{Shor1994} and simulations of quantum systems~\cite{Lloyd1996}.
The development of quantum computing during the past four decades~\cite{Fedorov2022}, both in theory and experiments, has made it possible to enter the so-called noisy intermediate-scale quantum (NISQ) era~\cite{Preskill2018}, 
which is manifested by the appearance of non-fault-tolerant quantum devices of a limited scale. 
Despite their restricted capabilities, NISQ devices already compete with and arguably even outperform the most powerful classical computers~\cite{Martinis2019,Pan2020, Pan2021-4,Pan2021-5,Morvan2023} in solving certain specific tasks.
Further progress in the development of quantum computing devices requires scaling the size of a quantum computing platform without degrading the quality of the control, which is an outstanding challenge. 

A computing paradigm that is at the foundation of most existing NISQ devices is the quantum digital circuit, also known as the gate-based model~\cite{Brassard1998,Ladd2010,Fedorov2022}. 
In a pioneering work,~\cite{DiVincenzo2000} formulated five qualitative requirements, known as DiVincenzo's criteria, that a physical setup must satisfy in order to support gate-based quantum computing. 
The first criterion, which is well-known, is that scalable physical systems with well-characterized two-level quantum subsystems, known as quantum bits or qubits, are used. 

Together with an accurate knowledge of the physical parameters of the considered system (including the internal Hamiltonian and interactions with other systems), {\it well characterized} also implies that a qubit, which can be realized via an arbitrary, possibly multilevel, quantum object, almost always remains in the subspace of these two levels.
This is a relevant argument since any leakage of quantum information can then be regarded as an error.
However, this is primarily true for situations where the transfer of information encoded in qubit states to higher energy levels is beyond the experimental control of the system~\cite{Lloyd1993}.
Alternatively, if one can control more than two levels, then one can extend the first criteria to the need for ``a scalable physical system with well-characterized, controlled {\it qudits}'' -- $d$-level quantum objects with $d>2$.
Here the phrase {\it well-characterized and controlled}, again indicates that the system almost always remains in the subspace of $d$ levels that are under control.
The other four DiVinchenzo criteria, which are related to state initialization, long enough coherence times, applying a universal set of operations, and making readout measurements, are generalized to the qudit case in a straightforward manner.

The aforementioned observation has become a driving force in studies of qudit-based quantum computing~\cite{Sanders2020}.
First ideas in this domain include multivalued logic gates for quantum computation, allowing for reduction in the required number of quantum information carries to realize a quantum algorithm~\cite{Muthukrishnan2000}, 
encoding multiple qubits in a single qudit~\cite{Kessel1999,Kessel2000,Kessel2002,Kiktenko2015,Kiktenko20152, shivam2024, Cao2023}, and, more recently, the use of qudits for more efficient decompositions of multiqubit gates, 
such as the Toffoli gate~\cite{Ralph2007,Wallraff2012,Kiktenko2020}.
Notably, the first realization of controlled-NOT logic gate involved a case in which two quantum bits were stored in the internal and external degrees of freedom of a single trapped atom~\cite{Wineland1995},
i.e., in a system of two qubits encoded in a single qudit.
Other branches of studies are related to measurement-based quantum computing models~\cite{Zhou2003,Raussendorf2011} and quantum error correction with qudits~\cite{Campbell2014,Svore2017}.
These ideas have been supported by deep analyses of entanglement structures of qudits, which are much more rich than qubits~\cite{Berry2002,Zeilinger2011,Song2016}, and enhancement of experimental control over qudits,
which led to demonstrations of universal multiqudit processors based on various physical platforms, 
such as 
superconducting circuits~\cite{Martinis2009,Wallraff2012,Gustavsson2015,Ustinov2015,Hill2021,Blok2021,Roy2023,Kazmina2024demonstration}, 
quantum light~\cite{White2009,Morandotti2017,Zeilinger2018,Wang2022}, 
trapped ions~\cite{Low2020,Ringbauer2021,Kolachevsky2022,Kazmina2024demonstration,Hrmo2023},
molecular magnets~\cite{Balestro2017,Chiesa2023},
and color centers~\cite{Fu2022}.
Many other physical systems are considered promising platforms for qudits ~\cite{Sawant2020,Weggemans2022}.
To some extent, qudits are already used in certain qubit-based quantum processors. 
For example, in superconducting and neutral-atom-based platforms, a two-qubit ${\sf CZ}$ gate can be realized using additional levels: 
it is either the third level of superconducting anharmonic oscillators~\cite{Oliver2019,Hill2021} or Rydberg states of atoms~\cite{Lukin2000,Saffman2022,Sanders2023}.
From this perspective, these systems can be partially considered as qutrits: three-level quantum systems.

This Colloquium is devoted to leveraging qudit-based hardware platforms, which operate with qudits, for running qubit-based algorithms (i.e., ones that are designed to be realized with a set of two-level systems).
This approach contrasts with a straightforward way of considering qudit-based computing as a multivalued generalization of the qubit-based paradigm --- a converting of a ``numeral base'' from 2 to $d$.
Acknowledging the importance of a multivalued approach, which was discussed in detail by~\cite{Sanders2020}, here we focus on perhaps more implicit advantages, 
which the use of additional levels is able to bring to the domain of quantum computing, namely, simplification of multiqubit gate realization and storage of a number of qubits in the state space of a single qudit.
These are of particular importance for the NISQ era since all the existing physical platforms face the scalability challenge. 

The idea of qudit-based information processing has also been intensively studied in various domains of quantum information science and technology, 
specifically, studies of fundamental aspects of quantum physics~\cite{Zeilinger2011,Fu2022}, in particular, 
tests involving Bell inequalities~\cite{Pironio2017, Dada2011, Gisin2002-2,Brunner2010},
as well as in the realizations of quantum teleportation~\cite{Hu2020, Pan2019}, 
quantum key distribution~\cite{Zheng2023, Tittel2000,Bourennane2001,Gisin2002,Sych2004,Mirhosseini2015,Gauthier2017,Karimi2017}, 
and quantum sensing~\cite{Blatter2018, Weides2018, Kristen2020}. 
However, these topics are beyond the scope of this Colloquium.

\begin{figure*}
\centering
\includegraphics[width=\linewidth]{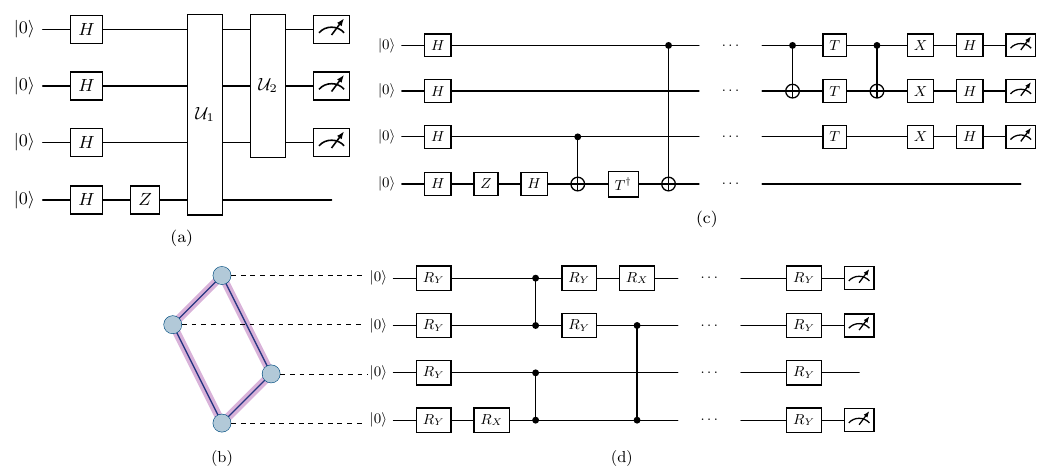}
\caption{(a) Example of a high-level quantum circuit, involving standard Hadamard and readout operations together with the complex multiqubit operations $\mathcal{U}_1$ and $\mathcal{U}_2$.
    (b) Architecture of a quantum processor with a restricted coupling map that is related to the making of the entangling gates.
    (c) Schematic results of a transpilation of the circuit from (a) in two- and single-qubit gates.
    (d) The resulting circuit ready for running on the real processor (b).
    $R_Y$ and $R_X$ denote single-qubit gates, and a standard notation for the {\sf CZ} gate is used.
    }
    \label{fig:qubit_circs_example}
\end{figure*}

The Colloquium is organized as follows.
In Sec.~\ref{sec:gate-based_model_of_qc} we review the standard circuit model of quantum computing with qubits and then generalize it to the qudit case.
In Sec.~\ref{sec:qudit-assisted_decs} we consider techniques of assisting upper levels of qudits for simplifying the implementation of multiqubit gates.
In Sec.~\ref{sec:embedding_several_qubits_in_qudit} we discuss the use of single qudits for embedding states of multiple qubits and related issues of making multiqubit gates, which couple qubits embedded in distinct qudits.
In Sec.~\ref{sec:exp_progress} we review recent experimental results in the development of qudit-based hardware.
We summarize, discuss open problems that are crucial for further progress in qudit-based computing, and make concluding remarks in Sec.~\ref{sec:concl}.

\section{Circuit model of quantum computing} \label{sec:gate-based_model_of_qc}

\subsection{Gate-based quantum computing with qubits}

We start with an overview of the circuit model of quantum computing with qubits.
For simplicity, we consider a non-fault-tolerant scenario, which typically takes place in the current NISQ era, where quantum error correction is not employed (i.e., we consider physical qubits). 

In the gate-based model, a quantum computer operates by executing a series of instructions (formerly known as a circuit) that are related to the manipulation of a set of qubits.
A quantum circuit can be formulated at different levels of abstraction.
As with a classical program, which can be expressed using high-level programming languages like {\sf C} and {\sf Python}, the high-level representation of a quantum circuit may involve complex instructions such as queries to quantum oracles, the application of the quantum Fourier transform, and multiqubit phase inversion (or diffusion) operations; see the example in Fig.~\ref{fig:qubit_circs_example}(a).
This is the usual approach for formulating and developing quantum algorithms. 

To run a high-level circuit on real hardware, the circuit has to be compiled, i.e., decomposed into elementary, also known as {native}, instructions comprehensible by a particular architecture of quantum processors.
Typically, the set of native instructions consists of the following operations: (i) initialization of a qubit in the state $\ket{0}$; (ii) creation of continuously parametrized (up to certain precision) single-qubit operations, for example,
\begin{equation}
    R_{\phi}(\theta)=\exp(-\imath\sigma_\phi\theta/2),
\end{equation}
where $\sigma_\phi := \sigma_{X}\cos\phi + \sigma_{Y}\sin\phi$ and $\imath$ represents the imaginary unit $\imath^2=-1$; 
(iii) formulation of some fixed entangling two-qubit gate, for example, the controlled phase
\begin{equation} \label{eq:CZ}
    {\sf CZ}: \ket{x,y} \mapsto (-1)^{xy}\ket{x,y}, \quad x,y\in\{0,1\},
\end{equation}
or the imaginary SWAP (iSWAP)
\begin{equation}
    {\sf iSWAP}: \ket{x,y} \mapsto \imath^{x\oplus y}\ket{y,x}, \quad x,y\in\{0,1\}
\end{equation}
gates; and (iv) performing a single-qubit computational basis ($\sigma_{Z}$) readout measurement.
Hereinafter, $\ket{0}$ and $\ket{1}$ denote qubits' computational basis states, 
\begin{equation}
    \sigma_{X} = \begin{pmatrix}0&1\\1&0\end{pmatrix},\quad 
    \sigma_{Y} = \begin{pmatrix}0&-\imath\\\imath&0\end{pmatrix},\quad
    \sigma_{Z} =
    \begin{pmatrix}1&0\\0&-1\end{pmatrix}
\end{equation}
are standard Pauli matrices, and a circled plus stands for a modulo 2 addition.
Entangling gates can usually be applied within a restricted coupling map representing the connectivity between qubits within the physical system; see Fig.~\ref{fig:qubit_circs_example}(b). 

The compilation can be conceptually divided into two major steps: (i) transpilation and (ii) final compilation.
The goal of transpilation is to decompose all necessary multiqubit operations into arbitrary single-qubit gates and a fixed two-qubit gate.
Typically, the controlled NOT (CNOT)
\begin{equation}
    {\sf CX}: \ket{x,y} \mapsto \ket{x,y\oplus x}, \quad x,y\in\{0,1\},
\end{equation}
two-qubit gate is used; see Fig.~\ref{fig:qubit_circs_example}(c).
The goal of the final compilation is to decompose the input circuit into native operations specific to the particular processor.

\begin{figure}
\centering
\includegraphics[width=\linewidth]{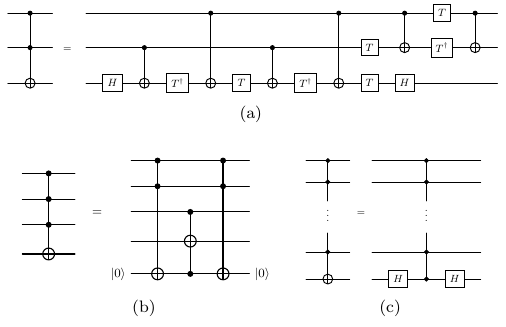}
\caption{(a) A standard decomposition of three-qubit Toffoli gate ${\sf C}^2{\sf X}$ into a sequence of controlled NOT (${\sf CX}$) gates.
Standard notations for controlled NOT, (generalized) Toffoli, Hadamard (${H}$), and $\pi/8$ ({T}) gates are used.
Note that ${\sf CX}$ gates are applied within an all-to-all coupling map.
(b) The idea behind an ancilla-qubit-based decomposition of a generalized Toffoli gate using an example of a ${\sf C}^3{\sf X}$ decomposition in three ${\sf C}^2{\sf X}$ gates.
(c) The idea behind switching between ${\sf C}^{N-1}{\sf X}$ and ${\sf C}^{N-1}{\sf Z}$ gates.
}
\label{fig:Toffoli_qubits}
\end{figure}

One of the most important practical tasks arising in the transpilation step is a decomposition of Toffoli
\begin{equation} \label{eq:Toffoli}
    {\sf CCX}\equiv {\sf C}^2{\sf X}: \ket{x_1,x_2,y} \mapsto \ket{x_1,x_2,y \oplus x_1x_2}
\end{equation}
and generalized $L$-controlled Toffoli 
\begin{equation} \label{eq:gen-Toffoli}
    {\sf C}^{L}{\sf X}: \ket{x_1,\ldots,x_L,y} \mapsto \ket{x_1,\ldots,x_L,y\oplus (x_1\ldots x_L)}
\end{equation}
gates.
These operations appear in oracles of classical Boolean functions.
They are used in Shor's algorithm~\cite{Shor1994, Antipov2022},
play a central role in the diffusion operator in Grover's algorithm~\cite{Grover1997}, and are also employed for decomposing an arbitrary multiqubit gate~\cite{NielsenChuang2000}.
In Fig.~\ref{fig:Toffoli_qubits}(a), we illustrate a standard transpilation of ${\sf C}^2{\sf X}$ gate using six entangling ${\sf CX}$ gates and several single-qubit gates.
It is known how to decompose ${\sf C}^{L}{\sf X}$ gate for $L>2$ acting on $L+1$ qubits with ${\cal O}(L^2)$ entangling gates~\cite{Barenco1995}. 
However, the decomposition can be significantly simplified down to ${\cal O}(L)$ entangling gates using ancillary qubits (also called ancillas).
In Fig.~\ref{fig:Toffoli_qubits}(b) we show an example of how to reduce a decomposition of ${\sf C}^{3}{\sf X}$ to three ${\sf C}^{2}{\sf X}$ gates by leveraging a ``clean'' ancilla -- an extra qubit prepared in the known state $\ket{0}$.
One can think of ${\sf C}^{L}{\sf X}$ as a kind of quantum program, while the employed ancillas correspond to using an extra computational space to run this program.
We note that according to the general paradigm of quantum computing, the program has to be reversible, so the initial state of all ancillas has to be restored.
Here we also note that a generalized Toffoli gate ${\sf C}^{L}{\sf X}$ can be obtained from a ``symmetric'' multicontrolled-phase gate
\begin{equation} \label{eq:gen-CZ}
    {\sf C}^{L}{\sf Z}: \ket{x_1,\ldots,x_{L+1}} \mapsto (-1)^{x_1\ldots x_{L+1}}\ket{x_1,\ldots,x_{L+1}},
\end{equation}
by specifying the target qubit with Hadamard gates; see Fig.~\ref{fig:Toffoli_qubits}(c).
In this way, the tasks of decomposing ${\sf C}^{L}{\sf Z}$ and ${\sf C}^{L}{\sf X}$ gates are equivalent from a practical point of view.

After the high-level circuit is transpiled down to single- and two-qubit gates, it can be supplied as input for the final compilation step; see Fig.~\ref{fig:qubit_circs_example}(d).
Typically, the architecture of the processor provides a simple way to transform ${\sf CX}$ gates into native two-qubit gates with the assistance of extra single-qubit operations; see ~\cite{Maslov2016}.
At the same time, there is a direct way of making an arbitrary single-qubit gates given the possibility of making Bloch rotations around two orthogonal axes (for example, $R_X(\theta):=R_{\phi=0}(\theta)$ and $R_Y(\theta):=R_{\phi=\pi/2}(\theta)$).
The main remaining problem is to adjust the pattern of two-qubit gates of the input circuit with the coupling map of the processor.
From a theoretical point of view, it can always be done in a straightforward manner by adding a necessary number of ${\sf SWAP}: \ket{x,y}\mapsto \ket{y,x}$ gates, which brings together qubits that have to be coupled.
Note that a single {\sf SWAP} gate can be realized with three {\sf CX} gates~\cite{NielsenChuang2000}, so they can also be decomposed to native single- and two-qubit gates. 
However, from a practical point of view, it is highly preferable to consider additional optimizations,
for example, by combining single-qubit gates, or optimizations related to the possibility of reassigning qubits of the input circuit with qubits of the processor during realization of the circuit 
(after bringing qubits together, it is not mandatory to move them back exactly to their starting positions).
We note that some of these kinds of optimization are already included in the functionality of modern packages for programming quantum circuits, for example, {\sf qiskit}~\cite{Qiskit2023} or {\sf cirq}~\cite{cirq2023}.
After the final transpilation step, the circuit is ready to run running on real hardware.

\subsection{Generalization to qudits} \label{sec:qudit-model}

The concept of qudit-based computing follows the idea of qubit-based computing with the only difference being that quantum information carriers are of a dimension $d>2$.
In this Colloquium, we use the term {\it qudit} to refer to a physical system with $d$ orthogonal quantum states that can be manipulated for various purposes of quantum information processing. 
Note that the $d$ states of a qudit are not necessarily all used simultaneously as computational basis states. 
As we later discuss, it can be advantageous to use a subset of the states of a qudit directly for computation while using the remaining states as auxiliary states for simplifying the implementation of complex gates.

At the hardware level, a qudit-based processor is characterized by a set of native instructions related to operations with multilevel objects.
For existing qudit-based processors, this set typically includes an initialization of qudits in the state $\ket{0}$, the performance of single-qudit and two-qudit operations, and the completion of readout measurement in the computational basis $\ket{0}, \ket{1}, \ldots, \ket{d-1}$.
The measurement can be realized in a ``single-shot'' manner or by making a sequence of quantum nondemolition measurements of the form $\{\ket{x}\bra{x}, \mathbb{1}-\ket{x}\bra{x}\}$, where $x=0,\ldots,d-1$ and $\mathbb{1}$ stands for $d\times d$ identity.

\begin{figure}
    \centering
    \includegraphics[width=\linewidth]{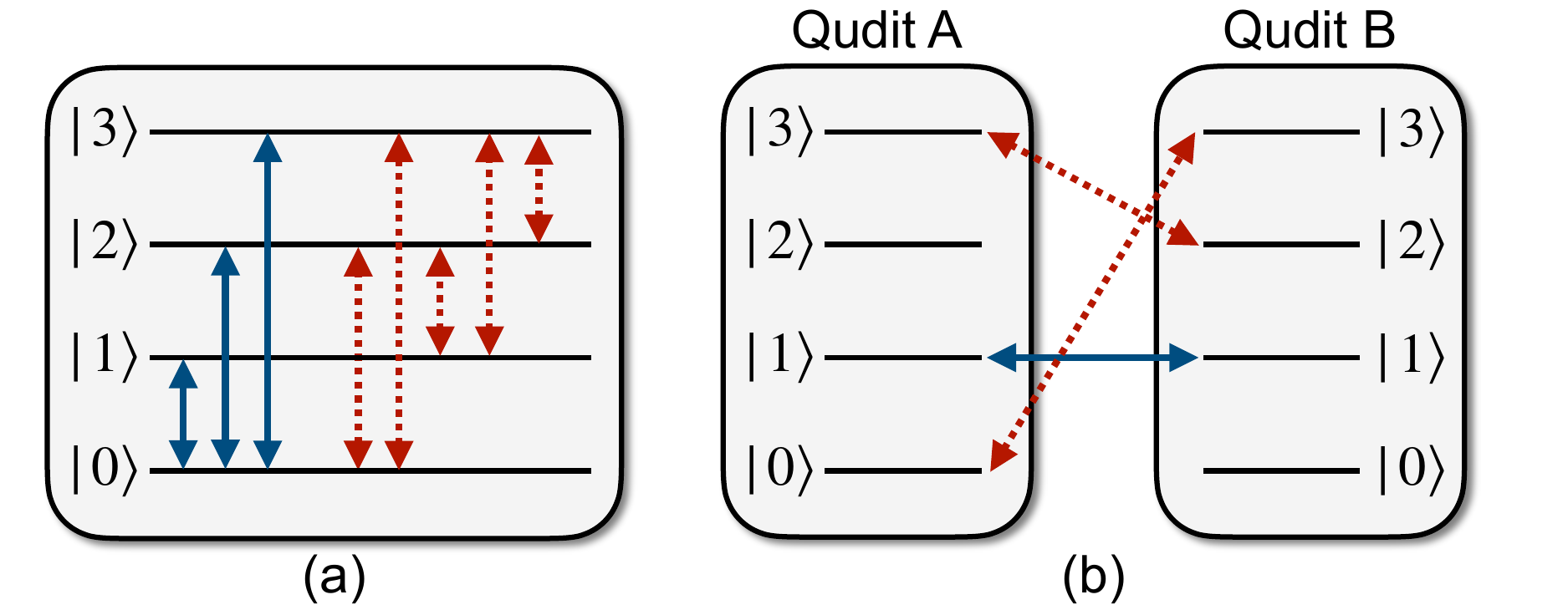}
    \caption{(a) Illustration of single-qudit gates.
    The solid arrows correspond to the allowed transitions between levels.
    The dashed arrows show the transitions accessible by combining operations on allowed transitions[for example, $R^{12}_\phi(\theta)=
        R_{Y}^{(01)}(\pi)
    	R_\phi^{(02)}(\theta)
    	R_{Y}^{(01)}(-\pi)$].
    (b) Schematic of a two-qudit ${\sf CPh}^{1|1}$ gate (solid arrow). 
    Other operations such as ${\sf CPh}^{0|3}$ and ${\sf CPh}^{3|1}$ (the dashed arrows) can be obtained by surrounding ${\sf CPh}^{1|1}$ with single-qudit gates [for example, ${\sf CPh}^{0|3}= \left(R_Y^{01}(-\pi)\otimes R_Y^{13}(-\pi)\right) {\sf CPh}^{1|1} \left(R_Y^{01}(\pi)\otimes R_Y^{13}(\pi)\right)$]. }
    \label{fig:qudit_natives}
\end{figure}

Single-qubit operations typically have the form 
\begin{equation} \label{eq:Rij}
    R_{\phi}^{ij}(\theta)=\exp(-\imath\sigma_\phi^{ij}\theta/2),
\end{equation}
where {$\sigma_\phi^{ij} := e^{-\imath\phi}\ket{i}\bra{j} +  
    e^{\imath\phi}\ket{j}\bra{i}$},
and pairs $(i,j)$ are taken from the set of allowed transitions between the qudit levels.
The parameter $\phi$ can typically be controlled by the phase of a physical pulse applied, whereas $\theta$ can be controlled by the pulse's duration and/or intensity.
Given a connected graph of allowed transitions, an operation $R_{\phi}^{ij}(\theta)$ at any transition $(i,j)$ can be realized by combining operations from the set of allowed ones~\cite{Mato2022, Blok2021}; see Fig.~\ref{fig:qudit_natives}(a).
In what follows we fix the notations 
\begin{equation}
	R_{X}^{ij}(\theta):=R_{\phi=0}^{ij}(\theta), \quad R_{Y}^{ij}(\theta):=R_{\phi=\pi/2}^{ij}(\theta).
\end{equation}
We also introduce a single-qudit phase operation
\begin{equation} \label{eq:Ph_i}
    {\sf Ph}^{i}(\theta): \begin{cases}
        \ket{x} \mapsto e^{\imath\theta} \ket{x},\quad&\text{if~}x=i,\\
        \ket{x} \mapsto \ket{x},\quad&\text{otherwise.}
    \end{cases}
\end{equation}
Usually, these gates can be realized virtually by commuting them to the beginning or the end of the circuit (where these gates can be discarded) and updating phases of physically applied pulses during each commutation [for example, $R^{ij}_\phi(\theta){\sf Ph}^i(\theta')={\sf Ph}^i(\theta')R^{ij}_{\phi+\theta'}(\theta)$ and $R^{ij}_\phi(\theta){\sf Ph}^k(\theta')={\sf Ph}^i(\theta')R^{ij}_{\phi}(\theta)$ for $k\neq i~\text{or}~j$).
See~\cite{mckay2017efficient, Blok2021} for a more detailed discussion and relevant examples.

As a basic two-qudit gate, we consider the following three options: the qudit controlled-phase gate
\begin{equation} \label{eq:CPh_qd}
    {\sf CPh}^{i|j}:
    \begin{cases}
        &\ket{i,j} \mapsto -\ket{i,j}, \\
	&\ket{x,y} \mapsto \ket{x,y}\quad  x\neq i\text{~or~}y\neq{j}, \\ 
    \end{cases}
\end{equation}
the qudit iSWAP gate 
\begin{equation} \label{eq:iSWAP_qd}
    {\sf iSWAP}^{ij|kl}(\theta):
    \begin{cases}
        \ket{i,k} \mapsto {e^{\imath\theta}}\ket{j,l}, \\
        \ket{j,l} \mapsto {e^{\imath\theta}}\ket{i,k},
    \end{cases}
\end{equation}
and the qudit M$\o$lmer-S$\o$rensen, which is widely used in trapped-ion quantum computing~\cite{Molmer-Sorensen1999,Molmer-Sorensen1999-2,Molmer-Sorensen2000},
\begin{equation} \label{eq:XX_qd}
    {\sf XX}^{ij|kl}_{\phi,\theta}(\chi) = \exp(-\imath \sigma^{ij}_{\phi}\otimes \sigma^{kl}_{\theta} \chi).
\end{equation}
For later convenience, we fix the notation ${\sf XX}^{ij|kl}(\chi):={\sf XX}^{ij|kl}_{\phi=0,\theta=0}(\chi)$.
{In addition, if the affected levels are unspecified, we use the bullet symbol instead: ${\sf XX}^\bullet$, ${\sf CPh}^\bullet$, etc.}

Entangling gates of the form~Eq.~\eqref{eq:CPh_qd}-\eqref{eq:XX_qd} naturally appear in various qudit processors~\cite{Hill2021,Roy2023,Ringbauer2021,Kolachevsky2022,Fischer2023,Senko2020}.
We note that from a practical point of view, it can be preferable to fix the affected levels, which are specified by the parameters $i,j,k,$ and $l$, for a native two-qudit gate 
and then construct the desired two-qudit operations, acting on arbitrary levels, by surrounding the native one with single-qudit operations; see Fig.~\ref{fig:qudit_natives}(b).
In this way, the same interaction that is used for realizing a two-qubit gate for a particular platform can be leveraged within the qudit-based approach: to directly operate with the upper levels, it is enough to use single-qudit gates only.

{From the qubit-based perspective, operations ${\sf CPh}^{1|1}$, ${\sf iSWAP}^{01|10}(\theta=\pm\pi/2)$, and ${\sf XX}^{01|01}(\chi=\pm \pi/4)$ generate a maximally entangled two-qubit state from an appropriate separable state.
Therefore, the cost of qudit-based ${\sf CPh}^{i|j}$ and ${\sf iSWAP}^{ij|kl}(\theta=\pm\pi/2)$ entangling gates from the viewpoint of making two-particle interactions is the same as in qubit-based schemes: the only difference is in making single-qudit operations affect the particular levels specified by $i,j,k,$ and $l$.
The situation with ${\sf XX}^\bullet(\chi)$ gates is slightly more complicated.
In qudit-based implementations, we further consider $\chi=\kappa \pi/4$, with $\kappa$  a positive integer. 
On the one hand, a direct way to compare qubit-based and qudit-based architectures is to view ${\sf XX}^{ij|kl}(\kappa\pi/4)$ as a sequence of $\kappa$ ${\sf XX}^{ij|kl} (\pi / 4)$ operations.
On the other hand, the results reported by~\cite{Moses2023} show that the infidelity of Mølmer-Sørensen (MS) gate as a function of  the input argument $\chi$  can be approximately described by $C_1+\chi C_2$ with the comparable positive constants $C_1$ and $C_2$. 
This implies that ${\sf XX}^{ij|kl}(\kappa\pi/4) $ may be more advantageous than simply a $\kappa$-fold repetition of ${\sf XX}^{ij|kl}(\pi/4)$ [since $\kappa (C_1 + C_2\pi/4) > C_1 + \kappa C_2\pi/4$]. 
Furthermore, recent experimental findings, reported by~\cite{nikolaeva2024ions} demonstrated the superiority of a Toffoli gate implemented using three ${\sf XX}^{01|01}(\pi/2)$ gates over the conventional decomposition using six ${\sf XX}^{01|01}(\pi/4)$ operations.
To account for potential improvements from combining ${\sf XX}^\bullet(\chi)$ gates, we provide both the number of ${\sf XX}^{ij|kl}(\kappa \pi/4)$ gates with $\kappa$ employed in a particular decomposition and the corresponding number of ${\sf XX}^{ij|kl}(\pi/4)$ gates when discussing ${\sf XX}^\bullet(\chi)$-based architectures.}

We note that by adding single-qudit operations, ${\sf CPh}^{i|j}$ can be transformed into a generalized controlled inversion gate
\begin{equation} \label{eq:CXijk}
    {\sf C X}^{i|jk}:
    \begin{cases}
        &\ket{i,j} \mapsto \ket{i,k}, \\
	&\ket{i,k} \mapsto \ket{i,j}, \\ 
	&\ket{x,y} \mapsto \ket{x,y},\quad \text{~if~} x\neq i, y\neq{j,k}
    \end{cases}
\end{equation}
considered, for example, in the qudit-based computations with photons~\cite{White2008, White2009, Wang2022}.
We also note that as with qubit-based architectures, the possibility to apply a two-qudit gate can be restricted by a partially connected graph of a coupling map.
Then an additional qudit SWAP gate may be required to compile a given circuit.

A straightforward way of using qudits is switching to a multivalued ($d$-ary) logic.
This approach relates to switching from the qubit Pauli group spanned by $\sigma_{X}$ and $\sigma_{Z}$ operators to the generalized Pauli group spanned ``shift'' and ``clock'' operators
\begin{equation}
	\sigma_{X}^{(d)}: \ket{x} \mapsto \ket{x+1~{\rm mod}~d}, \quad 
	\sigma_{Z}^{(d)}: \ket{x} \mapsto e^{\imath 2\pi x/d}\ket{x}.
\end{equation}
Using strings (tensor products) of $\sigma_{X}^{(d)}$ and $\sigma_{Z}^{(d)}$, it is possible to introduce generalized Clifford group and qudit-based stabilizer error correcting codes~\cite{Gottesman1997}.
Qudit versions of the seminal quantum algorithm, such as Grover's search and Shor's factorization, have been proposed~\cite{Saha2022, Parasa2011}.
For a more detailed discussion on qudit gate universality and qudit versions of the quantum algorithm, see the review by~\cite{Sanders2020}.

An alternative way of operating with qudits, the central one in this Colloquium is to employ qudit-based hardware for running qubit circuits.
The aim of rest of the Colloquium is to provide a comprehensive review of various techniques designed to improve the realization of qubit circuits by employing additional qudit space.
We pay special attention to how the type of realizable two-qudit gate~in Eq.~\eqref{eq:CPh_qd}, \eqref{eq:iSWAP_qd}, or~\eqref{eq:XX_qd} affects the applicability of these techniques.
Although the presented techniques have been developed for operating with physical qudits, the same methods can be formally applied to transpiling circuits, 
which operate with {\it logical qubits}, given that two-qudit logical operations of the same form are available.

\section{Qudit-assisted decomposition of multiqubit gates}
\label{sec:qudit-assisted_decs}

\begin{figure}
    \centering
    \includegraphics[width=0.6\linewidth]{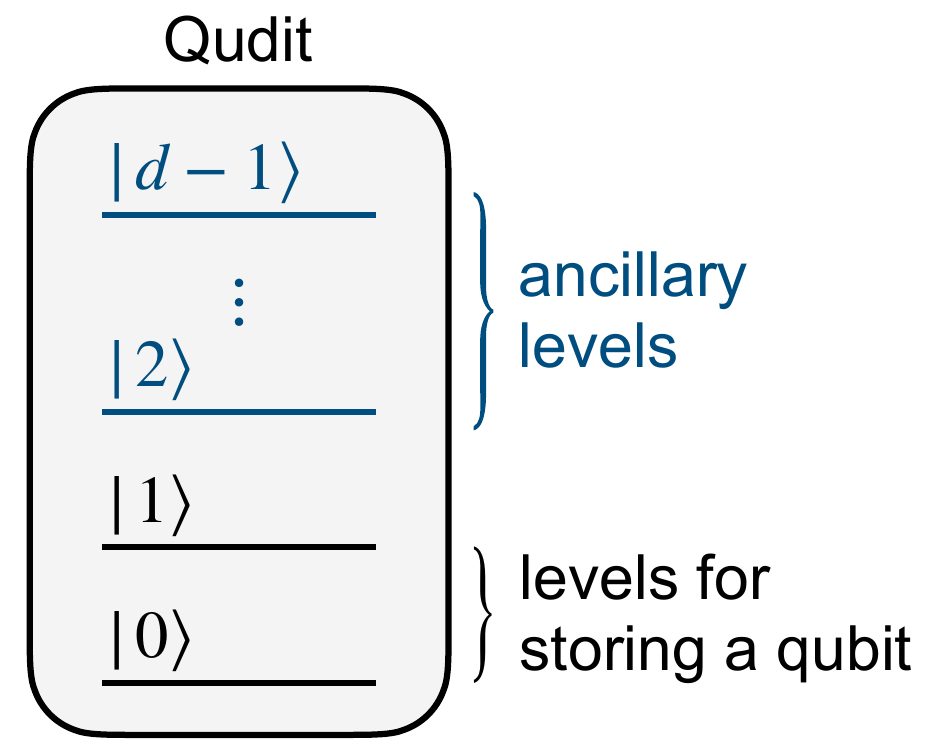}
    \caption{Structuring the computational space of qudits for simplifying decomposition of multicontrolled gates acting on a set of qubits.
    The first two levels are used for storing states of qubits involved in the decomposed gate.
    The upper $d-2$ ancillary levels serve as a temporary buffer and are populated only during implementation of the gate.}
    \label{fig:qudit_as_qubit_and_ancilla}
\end{figure}

Here we discuss the implementation of a multicontrolled-unitary gate using qudits.
In Sec.~\ref{sec:qubit-qubit-qutrit}-\ref{sec:multivalued} we focus on multicontrolled Toffoli gates, and in Sec.~\ref{sec:gen_to_CNU} we show how to generalize the introduced decompositions to the case of an arbitrary multicontrolled unitary.
Throughout this section a qudit state space is considered as a ``container'' for a qubit accompanied by $d-2$ ancillary levels, as shown in Fig.~\ref{fig:qudit_as_qubit_and_ancilla}.
In other words, the $d$-dimensional complex state space $\mathbb{C}^d$ of a qudit is decomposed as a direct sum, 
$\mathbb{C}^d = \mathbb{C}^2 \oplus \mathbb{C}^{d-2}$,
where the first two levels ($\mathbb{C}^2$) store the qubit's state and the remaining levels ($\mathbb{C}^{d-2}$) are ancillary.
We refer to the first two levels here as qubit levels of a qudit.

A qudit-based processor can then be viewed as an extension of the standard qubit-based device.
Given that the entire state of qudits of a qudit-based processor lies in the span of qubit states and that all applied gates, both local and entangling, fail to bring out population beyond this span, the processor effectively operates as a qubit-based one.
Therefore, to run a qubit circuit on this processor, one can compile the circuit down to gates that do not populate higher levels at all.
However, as we later show, the use of ancillary levels can significantly simplify the decomposition of multiqubit gates, especially the Toffoli in Eq.~\eqref{eq:Toffoli}, the generalized Toffoli~\eqref{eq:gen-Toffoli}, and multicontrolled-unitary gates.
The advantage is both in reducing the required number of entangling gates and removing the need for ancillary qubits.
In some sense, the ancillary levels in qudits substitute ancillary qubits in circuits like Fig.~\ref{fig:Toffoli_qubits}(b).
In the discussed qudit-assisted decomposition, ancillary levels are populated only during the implementation of multiqubit gates.
After making a gate acting on a state from qubits' subspace, all the population of ancillary levels is removed.
This is done to ensure that a valid unitary transformation is applied to the qubit subspace.

\subsection{Decomposition of the Toffoli gate with a single qutrit}
\label{sec:qubit-qubit-qutrit}

We consider the Toffoli gate in Eq.~\eqref{eq:Toffoli} acting in the qubits' subspaces of three particles.
Recall that the Toffoli gate can be decomposed using six two-qubit entangling gates acting within all-to-all topology, as shown in Fig.~\ref{fig:Toffoli_qubits}(a).
Now we demonstrate that if at least one of three particles has an ancillary level (i.e., is a qutrit), the Toffoli gate can be realized with three entangling gates applied within linear-chain topology.

\begin{figure}
\centering
\includegraphics[width=\linewidth]{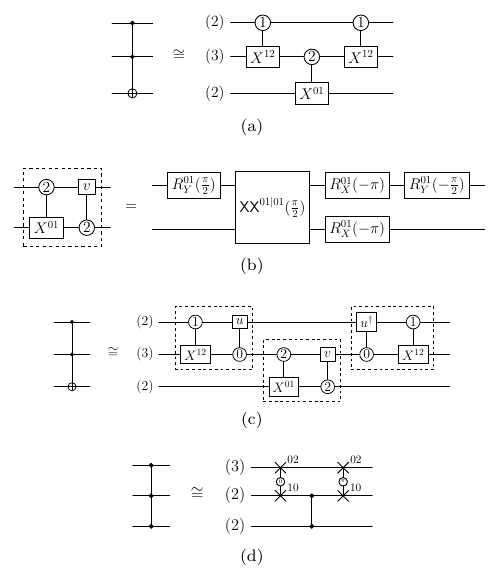}
\caption{Decomposition of the qubit Toffoli gate in the two-qubits-plus-qutrit system.
(a) Decomposition based on a generalized inversion ${\sf CX}^{i|jk}$ entangling gate.
(b) Depiction of the scheme for making a relative phase version of  ${\sf CX}^{2|01}$ gate out of ${\sf XX}^{01|01}(\frac{\pi}{2})$.
(c) Decomposition with relative phase generalized inversion gates.
In~(b) and~(c), $u$ and $v$ stand for some diagonal unitaries; see~\eqref{eq:Dstar}.
(d) Decomposition of ${\sf C}^2{\sf Z}$ (the Toffoli gate up to two Hadamard gates) using ${\sf iSWAP}^{02|10}(\theta)$ entangling gates. Note that the central controlled-phase gate can be realized as $[{\sf iSWAP}^{02|10}(0)]^2$.}
\label{fig:Toffoli_qutrits}
\end{figure}

In Fig.~\ref{fig:Toffoli_qutrits}(a), we present a decomposition based on generalized controlled inversion gates ${\sf CX}^{i|jk}$.
Hereinafter, a number in brackets to the left of a wire denotes the number of levels of corresponding particles occupied during a decomposition (this number can be less than the actual dimensionality of the qudit).
We use the approximately equals sign to stress the fact that the operation from the lhs is realized in the subspace of qubits' levels in the result of the operation on the rhs.
A ${\sf CX}^{i|jk}$ is depicted by $i$ in a circle on a control and $X^{jk}$ on a target.
Recall that we can also think about a ${\sf CX}^{i|jk}$-based decomposition as a decomposition based on ${\sf CPh}^{i|j}$ gates since ${\sf CX}^{i|jk}$ and ${\sf CPh}^{i|j}$ are converted into one another through the use of local operations.
The construction of the decomposition in Fig.~\ref{fig:Toffoli_qutrits}(a) is intuitive.
For the input state $\ket{x_1,x_2,y}$ ($x_1, x_2, y\in\{0,1\}$), the first ${\sf CX}^{1|12}$ gate populates an ancillary level $\ket{2}$ of the qutrit in the middle if and only if $x_1x_2=1$.
Then the central ${\sf CX}^{2|01}$ applies an inversion operation to the target qubit ($\ket{y}\mapsto\ket{y\oplus 1}$) for $x_1x_2=1$ and does nothing ($\ket{y}\mapsto\ket{y}$) for the other three possible values of $x_1$ and $x_2$.
The third ${\sf CX}^{2|01}$ gate restores (``uncomputes'') the initial state of the control particles.
A possible population of the ancillary level of the qutrit is removed, while the effective transformation in the qubit subspace becomes identical to the Toffoli gate in Eq.~\eqref{eq:Toffoli}.

A similar trick can be used for a qubit-qutrit-qubit system and the MS entangling gate in Eq.~\eqref{eq:XX_qd}.
By surrounding ${\sf XX}^{01|01}(\frac{\pi}{2})$ with single-qudit operations as shown in Fig.~\ref{fig:Toffoli_qutrits}(b), we arrive at the following two-qutrit unitary:
\begin{equation} \label{eq:phase-rel-CX}
    {\sf CX}^{2|01}_\star :=
    \imath\begin{pNiceMatrix}[first-row, last-col]
    00 & 01 & 02 & 10 & 11 & 12 & 20 & 21 &22&  \\
    {\bf 1}&&&&&&&&&00 \\
    &{\bf 1}&&&&&&&&01\\
    &&1      &&&&&&&02\\
    &&&{\bf 1}&&&&&&10\\
    &&&&{\bf 1}&&&&&11\\
    &&&&&-1&&&&12\\
    &&&&&&&{\bf 1}&&20\\
    &&&&&&{\bf 1}&&&21\\
    &&&&&&&&-\imath&22\\
\end{pNiceMatrix},
\end{equation}
where the numbers on the border represent the corresponding computational basis states and all unspecified entries are zero.
Equation~\eqref{eq:phase-rel-CX} is proportional to the product of ${\sf CX}^{2|01}$ and the diagonal relative phase unitary~\cite{Maslov2016}
\begin{equation} \label{eq:Dstar}
    {\sf D}_{\star} = \mathbb{1} \otimes (\ket{0}\bra{0} + \ket{1}\bra{1}) +
    \begin{pmatrix}
        1\\&-1\\&&-\imath 
    \end{pmatrix}
    \otimes \ket{2}\bra{2},
\end{equation}
which can be considered as a $\ket{2}$-controlled gate with a target (control) on the first (second) qutrits.
Note that the decomposition of ${\sf CX}^{2|01}_\star$ in Fig.~\ref{fig:Toffoli_qutrits}(b) does not include any operations on the upper qudit states $\ket{2}$; however, a closer examination reveals that this is in fact correct. 
The key point is that the ${\sf XX}^{01|01}(\frac{\pi}{2})$ gate acts as the identity on the states $\ket{2} \ket{\psi}$ and $\ket{\psi}\ket{2}$ for an arbitrary $\ket{\psi}$, while within the two-qubit subspace spanned by qubit states $\ket{x} \ket{y}$, with $x,y=0,$ and 1, it acts as $R_X^{01}(\pi)\otimes R_X^{01}(\pi)$.
We note that the control state $\ket{2}$ is in the orthogonal complement of the subspace spanned by $\ket{0}$ and $\ket{1}$, where the inversion made by ${\sf CX}^{2|01}$ takes place.
At the same time, within a qutrit-qubit subspace spanned by $\{\ket{x,y}:x=0,1,2,~y=0,1\}$  [the highlighted elements in~Eq.~\eqref{eq:phase-rel-CX}], 
${\sf CX}^{2|01}_\star$ acts exactly as ${\sf CX}^{2|01}$.
The resulting structure makes it possible to implement the Toffoli gate decomposition shown in Fig.~\ref{fig:Toffoli_qutrits}(c).
In that panel the first and third gates are conjugated phase-relative versions of ${\sf CX}^{1|12}$, which can be realized in a manner similar to ${\sf CX}^{2|01}_\star$ by replacing of affected levels using the surrounding with local inversion gates. 
The basic intuition behind Fig.~\ref{fig:Toffoli_qutrits}(c) remains the same as that behind Fig.~\ref{fig:Toffoli_qutrits}(a): the central inversion gate is activated by populating the ancillary state of the qutrit, which takes place only for the $\ket{1,1}$ initial state on the controls.
Note that relative phase diagonal unitaries coming from the first and third gates compensate each other, while the phase correction from the central gate acts beyond the involved qutrit-qubit subspace.

Finally, the generalized iSWAP gate-based decomposition of the Toffoli gate (more precisely, ${\sf C}^2{\sf Z}$) is shown in Fig.~\ref{fig:Toffoli_qutrits}(d).
As in Figs.~\ref{fig:Toffoli_qutrits}(a) and ~\ref{fig:Toffoli_qutrits}(c), it also consists of three entangling gates acting within a V-shaped pattern; however, it has some important conceptual differences.
Here the qutrit is located ``on the edge'' (the first particle), and the central operation is the two-qubit controlled-phase gate~in~Eq.~\eqref{eq:CZ}
(recall that ${\sf CZ}$ can be realized in the two-qubit subspace, such as by double application of ${\sf iSWAP}^{02|10}(0)$).
The idea of the first ${\sf iSWAP}^{02|10}(0)$ is to leave the population on $\ket{1}$ in the middle qubit only if the first control particle is also in $\ket{1}$.
Otherwise, the middle qubit goes to $\ket{0}$ and the central ${\sf CZ}$ does not bring any changes.
The central ${\sf CZ}$ also makes the identity operation if the middle qubit is initially in the state $\ket{0}$.
After uncomputation with ${\sf iSWAP}^{02|10}(\pi)={\sf iSWAP}^{02|10}(0)^\dagger$, which restores the population and phase changes that came with the first ${\sf iSWAP}^{02|10}(0)$, 
the entire state of the particles is returned to the qubit subspace with the acquired $-1$ phase only for the input basis state $\ket{1,1,1}$.
A decomposition of this type has been demonstrated experimentally in a superconducting platform~\cite{Wallraff2012}.

As we have seen, the schemes based on ${\sf CX}^{ij|k}$ and ${\sf XX}^{ij|kl}$ differ from the one based on ${\sf iSWAP}^{ij|kl}(\theta)$ 
in how the ancillary qutrit's state is employed to prepare for applying the central controlled-unitary gate -- the core of the realized Toffoli gate.
In the former case, the ancillary state serves an activator for the central gate, while in the later case, 
the ancillary level is used for absorbing the controlling state $\ket{1}$ for the central controlled gate if not all controls of the entire Toffoli gate are in the controlling state $\ket{1}$~\cite{Hanks2022}.
This difference plays a more prominent role in the generalization of introduced schemes.

\subsection{Decomposing Toffoli gates with an arbitrary number of controls within linear and starlike topologies}
\label{sec:linear_starlike}

The considered ideas can be used in decompositions of ${\sf C}^{N-1}{\sf X}$ (${\sf C}^{N-1}{\sf Z}$) gates with an arbitrary number of controls.
An important aspect of the extension is related to the connectivity topology.
Here we discuss two cases of the linear (nearest neighbors) and starlike ones.

\begin{figure}
\centering
\includegraphics[width=\linewidth]{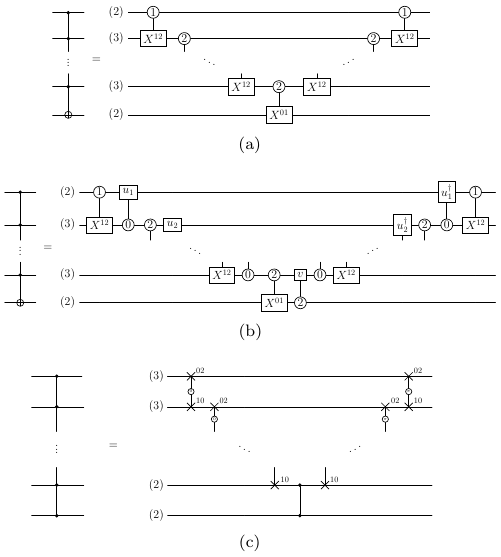}
\caption{Generalizing the ideas behind qutrit-based decomposition of the Toffoli gate to the case of ${\sf C}^{N-1}X$ (${\sf C}^{N-1}Z$) gates.
Qutrit-based decompositions using generalized inversion ${\sf CX}^{i|jk}$, phase-relative generalized inversion [constructed based on ${\sf XX}^{ij|kl}(\pi/2)$ according to Fig.~\ref{fig:Toffoli_qutrits}(c)], and ${\sf iSWAP}^{ij|kl}(\theta)$ gates are shown in (a), (b), and (c), respectively.
In (b), $u_i$ and $v$ stand for some diagonal unitaries that come from turning ${\sf XX}^{ij|kl}(\pi/2)$ into a phase-relative generalized inversion gate.}
\label{fig:Toffoli_linear}
\end{figure}

We correspondingly show the decompositions based on ${\sf CX}^{i|jk}$, ${\sf XX}^{ij|kl}(\pi/2)$, and ${\sf iSWAP}^{ij|kl}(\theta)$ gates acting within a linear-chain topology in Fig.~\ref{fig:Toffoli_linear}(a, b, c).
The resulting circuits have V shapes, involve a set of two qubits and $N-2$ qutrits, and contain $2N-3$ two-particle entangling operations.
As in Sec. \ref{sec:qubit-qubit-qutrit}, ancillary levels are employed either as controls for the central gate in the case of ${\sf CX}^{i|jk}$ and ${\sf XX}^{ij|kl}$ gates 
or as an absorber for deactivation of the controlling state $\ket{1}$ in the central gate in the case of ${\sf iSWAP}^{ij|kl}$ gates; see \cite{Zheng2012} for a discussion of ${\sf iSWAP}^{ij|kl}$-based scheme.
We also note that since gates on the left slope of the V-shaped circuit are conjugates to the one on the right slope, the additional optimization of removing single-qudit gates that face each other is possible.
{The experimental ${\sf XX}^{ij|kl}$-based decomposition of ${\sf C}^{N-1}X$ for $N\leq10$ on a trapped $^{171}{\rm Yb}^+$ ion platform was recently reported by~\cite{nikolaeva2024ions}.}

Given the random computational basis input state, the difference in the way that the ancillary levels are used yields the difference in the number of qutrits populated in the ancillary state before applying the central gate.
As an example, consider the input state $\ket{1,1,0,1,1,0,1,0}$.
By the time the central gate is applied, the state for the circuits in Fig.~\ref{fig:Toffoli_linear}(a,b) is $\propto\ket{1,{\bf 2},0,1,1,0,1,0}$: the number of $\ket{2}$s is limited by the first appearance of $\ket{0}$ in the input state.
At the same time, for the circuit from Fig.~\ref{fig:Toffoli_linear}(c), the state takes the form $\propto\ket{1,1,{\bf2,2}~0,{\bf 2},0,0}$.
The ancillary level is populated for every appearance of $\ket{0,1}$ during the computation (the appearance of ancillary states is highlighted for convenience).
As well studied by.~\cite{Hanks2022}, such behavior results in different resistivity to noise, 
taking into account that the upper ancillary level typically has fewer coherence times than qubit levels (from this point of view, the schemes based on controlled inversion are preferable to iSWAP-based ones).

\begin{figure}
\centering
\includegraphics[width=\linewidth]{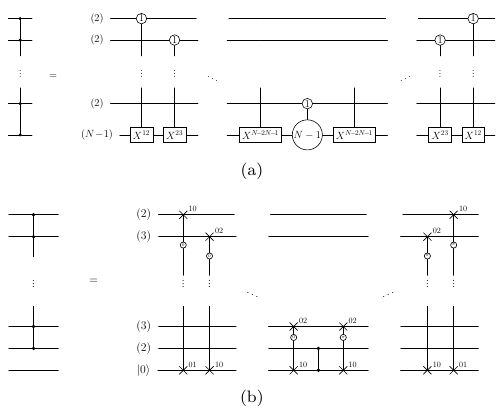}
\caption{Qudit-based decompositions of ${\sf C}^{N-1}{\sf Z}$ for quantum information carriers connected within a starlike topology.
(a) Decomposition based on generalized controlled inversion ${\sf CX}^{i|jk}$ gates acting on $N-1$ qubits coupled with an $N$-dimensional qudit. 
(b) iSWAP-based decomposition for $N-2$ qutrits and $3$ qubits.
Note that a clean ancillary qubit initialized in $\ket{0}$ is placed in the center of the star.}
\label{fig:Toffol_star}
\end{figure}

Another important special case of a coupling topology is the starlike one, where the multilevel qudit is connected to a number of qubits~\cite{Ralph2007}.
In Fig.~\ref{fig:Toffol_star} we show the decomposition of a ${\sf C}^{N-1}{\sf X}$ gate with an $N$-level qudit surrounded by $N-1$ qubits.
Hereinafter, ${\sf CPh}^{i|j}$ is depicted by connected circles with $i$ and $j$.
A similar decomposition was demonstrated experimentally using a photonic system~\cite{White2009}.
This scheme is especially suited for the ${\sf CX}^{i|jk}$ entangling gate.
The idea is that the central ${\sf CZ}^{1|N-1}$ gate adds a $-1$ phase factor if and only if the ladder of $N-2$ preceding the ${\sf CX}^{1|i,i+1}$ gates ($i=1,2,\ldots,N-2$) raises the state of the qudit up to $\ket{N-1}$; i.e., all qubits are in the state $\ket{1}$.
Then the mirror-symmetrical uncomputing sequence restores the initial state of controls.
As with the circuits in Fig.~\ref{fig:Toffoli_linear}, the resulting circuit consists of $2N-3$ entangling operations yet requires a high-dimensional $(N-1)$-level qudit.
This fact poses an obvious scalability limitation to this approach.

Another decomposition of ${\sf C}^{N-1}{\sf Z}$ realized for $N$ qutrits (or, more precisely, $N-2$ qutrits and $2$ qubits) coupled to the ancillary qubit within the starlike topology via ${\sf iSWAP}^{10|01}(\theta)$ and ${\sf iSWAP}^{02|10}(\theta)$ gates was considered both theoretically and experimentally by~\cite{Fang2023}; see the scheme in Fig.~\ref{fig:Toffol_star}(b).
Note that here the central operation is performed between the ancilla and one of the particles affected by the ${\sf C}^{N-1}{\sf Z}$ gate.
As experimentally demonstrated by~\cite{Fang2023}, 
this decomposition is particularly well suited to trapped-ion-based platform, where the central qubit can be represented by collective motional degree of freedom, surrounding qutrits (qubits) as inner states of trapped ions, and Cirac-Zoller entangling operation~\cite{CiracZoller1995}.

A similar system of $N$ qubits coupled to $N$-level ancillary qudit with the ability to make 
\begin{equation}\label{eq:cx-dary}
    {\sf C}\sigma_{X}^{(d)}: \ket{i}\ket{j}\mapsto \ket{i}\ket{(i+j)~{\rm mod}~d}
\end{equation}
entangling gates was considered by~\cite{Ionicioiu2009}.
It was shown that an $N$-qubit multicontrolled-unitary gate can be realized on qubits by first applying $N$ entangling gates between each qubit and the qudit and then by disentangling the qudit from qubits via measurement-based feedforward operation.
In some sense, the measurement-based feedforward trick is used to avoid an uncomputation sequence.
See~\cite{Ionicioiu2009} for more details.

\subsection{Generalization to an arbitrary topology}
\label{sec:arb_topol}

\begin{figure}
\centering
\includegraphics[width=\linewidth]{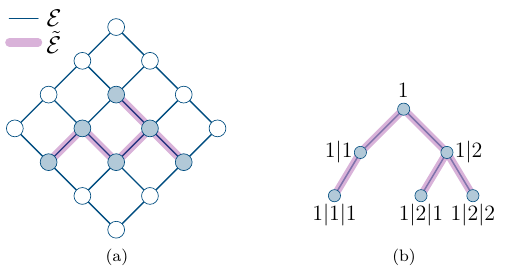}
\caption{(a) Example of a coupling graph ${\cal E}$ and a reduced cycle-free subgraph $\widetilde{{\cal E}}$ that connects particles affected by a ${\sf C}^{5}{\sf X}$ gate.
(b) The idea behind the notation of nodes in the tree $\widetilde{{\cal E}}$.
}
\label{fig:graph}
\end{figure}

We now consider the decomposition of ${\sf C}^{N-1}{\sf X}$ gate in the case of an arbitrary connectivity.
We introduce a graph ${\cal E}$  consisting of $N$ vertices, each of which corresponds to one of $N$ particles involved in ${\sf C}^{N-1}{\sf X}$, with edges defined by the possibility of realizing an entangling operation between corresponding particles.
We assume that ${\cal E}$ is a connected graph; otherwise, additional SWAP operations are required to make it connected.
We choose a cycle-free $N$-vertex subgraph (tree) $\widetilde{{\cal E}}$ inside ${\cal E}$; see an example in Fig.~\ref{fig:graph}(a).
A preferable way of choosing $\widetilde{{\cal E}}$ is such that the resulting tree has the lowest possible height.

Two schemes of implementing ${\sf C}^{N-1}{\sf X}$ gate with qudit-based entangling operations acting within $\widetilde{{\cal E}}$ are discussed here.
The first scheme, which was developed~\cite{Kiktenko2020}, is based on the ${\sf CX}^{i|jk}$ gate and requires the following relation for each qudit:
\begin{equation} \label{eq:d_and_k_condition}
    d_i\geq k_i+1,
\end{equation}
where $d_i$ is a dimension of the $i$th qudit ($i\in\{1,\ldots,N\}$) and $k_i$ is its number of connections to other qudits within $\widetilde{{\cal E}}$.
The second scheme is based on ${\sf iSWAP}^{ij|kl}$ and can be realized with qutrits only
(as shown later, two of the $N$ particles are allowed to have no ancillary levels, so effectively it is an $(N-2)$-qutrits-plus-two-qubits-based decomposition) \cite{Nikolaeva2022} .

Both schemes are constructed in such a way that adding gates to decomposition is accompanied by operations with the tree $\widetilde{{\cal E}}$ (from now on, we consider $\widetilde{{\cal E}}$ as an abstract mathematical object).
We introduce special notations for the vertices of $\widetilde{{\cal E}}$.
Let ${\bf 1}$ denote the root, ${\bf 1|1}, {\bf 1|2}, \ldots$ denote children of the root (nodes at the second layer), let ${\bf 1|a|b}$ denote the ${\bf b}$th child of ${\bf a}$th child of the root (the third level), etc.;  see also an example in Fig.~\ref{fig:graph}(b).
In addition, let ${\cal N}({\bf s})$ be the number of children of an arbitrary node ${\bf s}$.

\begin{figure}
\centering
\includegraphics[width=\linewidth]{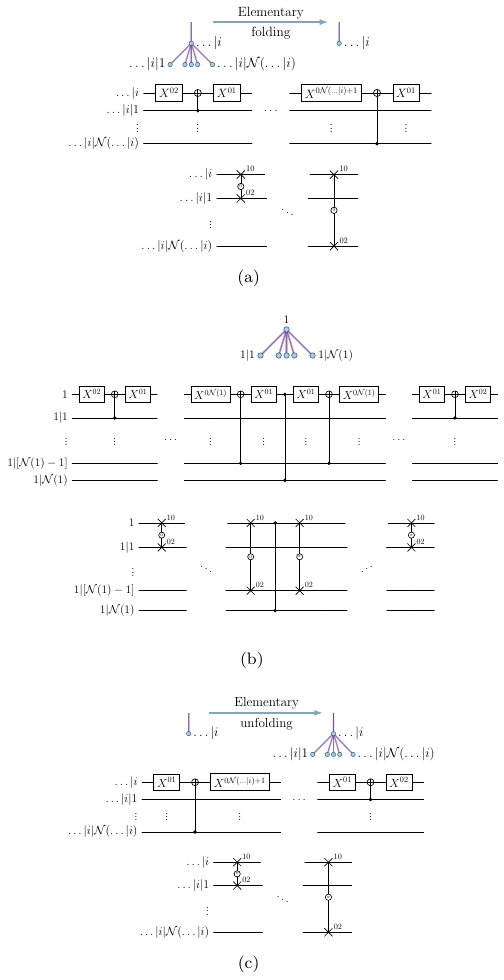}
\caption{Sequences of gates within the (a) folding, (b) central, and (c) unfoliding steps for schemes based upon ${\sf CX}^{i|jk}$ and ${\sf iSWAP}^{ij|kl}$. The gate $X^{0j}$ makes an inversion of populations at $\ket{0}$ and $\ket{j}$. 
For iSWAP-based folding and unfolding operations, the order of internal operations does not matter, since the third level of the child nodes is used.}
\label{fig:folding-unfolding}
\end{figure}

{Our decomposition process consists of three main steps: (i) folding, (ii) the central step, and (iii) unfolding.\footnote{These steps were also named (i) embedding, (ii) the controlled-unitary operation, and (iii) recovery by \cite{Chu2023}. To avoid   confusion in the definition of the term {\it embedding}, we use the terms {\it folding} and {\it unfolding} in the Colloquium.}
The folding step involves the sequential removal of leaves (nodes with no children) from $\widetilde{\cal E}$ until only the root node ${\bf 1}$ and its direct children ${\bf 1 | 1}, ..., {\bf 1} | {\cal N}(\bf 1)$ remain. 
The entire folding step is divided into so-called elementary folding operations, during which the leaves ${\bf s | 1},...,{\bf s} | {\cal N}({\bf s})$ are removed from a given node ${\bf s}$, making it a leaf itself. 
For example, an elementary folding operation for the tree shown in Fig.~\ref{fig:graph} would correspond to removing the nodes ${\bf 1|2|1}$ and ${\bf 1|2|2}$ and turning the node ${\bf 1|2}$ into a leaf.
Each elementary folding operation is accompanied by the addition of gates to qudits that correspond to nodes of ${\widetilde{{\cal E}}}$ participating in this folding operation.
The added sequences of gates within the ${\sf CX}^{\bullet}$ and ${\sf iSWAP}^{\bullet}$ schemes are shown in Fig.~\ref{fig:folding-unfolding}(a).
The purpose of the added gates is to ensure that the qudit ${\bf s}$ (or, more precisely, the qudit that corresponds to node ${\bf s}$ in $\widetilde{\cal E}$) is output in the state $\ket{1}$ if and only if all qudits ${\bf s}, {\bf s|1},\ldots,{\bf s}|{\cal N}({\bf s})$ were input in the state $\ket{1}^{\otimes ({\cal N}({\bf s})+1)}$.
Note that in the ${\sf CX}^{\bullet}$-based scheme, the number of required additional levels for ${\bf s}$ is equal to ${\cal N}({\bf s})-1$, which implies Eq.~\eqref{eq:d_and_k_condition}.
After completing all the elementary folding operations, the arrival of the root's children qudits ${\bf 1|1},\ldots,{\bf 1}|{\cal{N}}({\bf 1})$ to the all-1s state is equivalent to all-1s initial state of all the qudit leaves in the initial unfolded tree.
From the viewpoint of the graph in Fig.~\ref{fig:graph}(b), applying all the gates generated during the folding operations will result in the state $\ket{1}^{\otimes 2}$ on ${\bf 1|1}$ and ${\bf 1|2}$ if and only if all five qudits ${\bf 1|1|1}$, ${\bf 1|1}$, ${\bf 1|2|1}$, ${\bf 1|2|1}$, and ${\bf 1|2}$ were initially prepared in the $\ket{1}^{\otimes 5}$ state.
}

The approach for the central step is to apply the ${\sf C}^{{\cal N}({\bf 1})}{\sf Z}$ operation within the root and its children.
It is realized through the sequence of gates shown in Fig.~\ref{fig:folding-unfolding}(b).
We note that in the ${\sf iSWAP}^{ij|kl}$-based scheme, the third level of the root and one of its children are not explicitly involved (yet can be used to realize the operation ${\sf CZ}$).

The third unfolding step is a mirror reflection of the folding one.
Like the V-shaped schemes in Fig.~\ref{fig:Toffoli_linear}, it provides the uncomputation: 
states of particles are turned back to their initial ones and the possible population at ancillary levels is removed.
After completing all three steps, the only difference is that the $-1$ phase factor is acquired in the central step if and only if the particles were initially in the all-1s state.

\begin{figure}
    \centering
\includegraphics[width=\linewidth]{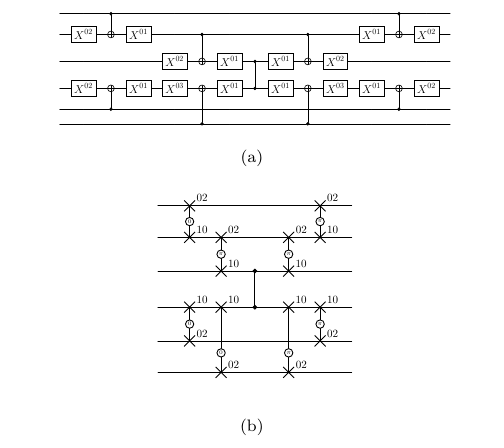}
    \caption{Explicit examples of ${\sf C}^5{\sf Z}$ decomposition with (a) ${\sf CX}^{i|jk}$ and (b) ${\sf iSWAP}^{ij|kl}$ entangling gates.
    The structure of the tree is taken from Fig.~\ref{fig:graph}(b).}
    \label{fig:arb-topology-example}
\end{figure}

We present explicit examples of ${\sf CX}^{i|jk}$- and ${\sf iSWAP}^{ij|kl}$-based decompositions in Fig.~\ref{fig:arb-topology-example}.
The resulting circuits in both schemes consist of $2N-3$ entangling gates. 
Their depth is determined by the topology of $\widetilde{\cal{E}}$, 
since elementary folding and unfolding operations can be done in parallel for different subtrees (this is why it is preferable to take an $\widetilde{\cal{E}}$ of the smallest possible height).

The presented schemes are more scalable than those considered previously.
For the ${\sf CX}^{i|jk}$-based scheme, the dimensionality of involved qudits is limited by the degree of nodes inside $\widetilde{\cal{E}}$, while for ${\sf iSWAP}^{ij|kl}$-based scheme, the qudit dimensionality $d=3$ is enough for any topology of connection. 
Note that the total number $N_{\rm anc}=N-2$ of the employed ancillary levels is the same in both schemes.
{We also note that the described ${\sf iSWAP}^{ij|kl}$-based $N$-qubit Toffoli gate decomposition was implemented experimentally on a superconducting qutrit processor for $N=4,6,8$ by \cite{Chu2023}.}

\subsection{Employing multivalued logic}
\label{sec:multivalued}

\begin{figure}
\centering
\includegraphics[width=\linewidth]{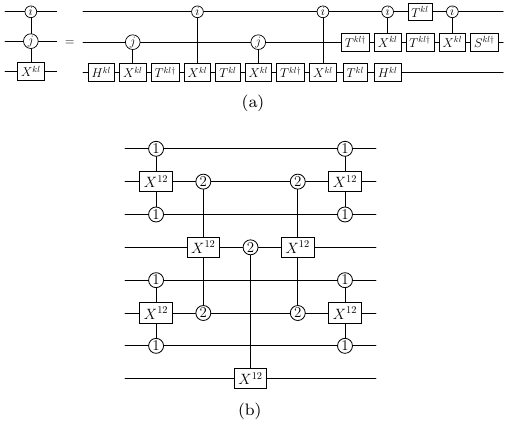}
\caption{(a) The decomposition of  the multivalued Toffoli gate ${\sf CCX}^{i|j|kl}$.
$U^{ij}$, with $U\in\{T,X,H\}$, denotes a standard gate $U$ acting in a two-level subspace spanned by $\ket{i}$ and $\ket{j}$.
Adapted from~\cite{di2011elementary}. 
(b) Decomposition of a ${\sf C}^{7}{\sf X}$ gate using multivalued Toffoli gates. Adapted from~\cite{Gokhale2019}.
}
\label{fig:multi-valued-USA}
\end{figure}

Another approach for decomposing multiqubit Toffoli gates, proposed~\cite{Gokhale2019} was based on borrowing results from multivalued (ternary) qutrit-based computations.
It employs the decomposition from~\cite{di2011elementary} of a ``fair'' ternary Toffoli ${\sf CCX}^{i|j|kl}$ gate, which swaps population between $\ket{k}$ and $\ket{l}$ for the target qutrit if the two control qutrits are in the state $\ket{i,j}$, doing nothing otherwise.
This decomposition, which is realized with ${\sf CX}^{i|jk}$ entangling operations, is shown in Fig.~\ref{fig:multi-valued-USA}(a).
Note that it mimics the one for the standard qubit Toffoli gate decomposition in Fig.~\ref{fig:Toffoli_qubits}(a).

The constructed ${\sf CCX}^{i|j|kl}$ gate can then be employed for making a ${\sf C}^{N-1}{\sf X}$ gate for $N=2^\kappa$ ($\kappa$ is a positive integer), as shown in Fig.~\ref{fig:multi-valued-USA}(b).
The main idea remains the same:
the central ${\sf CX}^{2|01}$ gate makes an inversion if all the controls were in $\ket{1,\ldots, 1}$ in the beginning.
The closing mirror-symmetrical part removes the possible populations of $\ket{2}$ and finishes the decomposition.
Compared to the previously considered ${\sf CX}^{i|jk}$-based decomposition for qutrits in a chain (see Fig.~\ref{fig:Toffoli_linear}), 
the presented one achieves a logarithmic depth at the cost of an increase in the number of entangling operations from $2N-3$ to $6N-5$.
The threefold increase comes with ternary Toffoli gates.
Note that since the decomposition of the ternary Toffoli gate [see Fig.~\ref{fig:multi-valued-USA}(a)] requires all-to-all connectivity, the entire decomposition in Fig.~\ref{fig:multi-valued-USA}(b) is realized over a triangulum-based binary tree.

The resulting characteristics of all the introduced decompositions of ${\sf C}^{N-1}{\sf X}$ gate are summarized in Table~\ref{tab:comparison}.

\begin{table*}[]
    \centering
    \begin{tabular}{c|c|c|c|c|c}
         Entangling gate & Qudit dimension & Coupling map & No. of ent. gates & Circuit depth & Reference \\ \hline\hline
         ${\sf CX}^\bullet$ & $d=N$ & star
         & $2N-3$ & ${\cal O}(N)$ & \cite{White2009}\\
          ${\sf CX}^\bullet$ & $d=3$ & linear
         & $2N-3$ & ${\cal O}(N)$ & \cite{Wallraff2012}\\ 
         ${\sf CX}^\bullet$ & $d=3$ & triangulum-based tree
         & $6N-5$ & ${\cal O}(\log N)$ & \cite{Gokhale2019}\\
         ${\sf CX}^\bullet$ & $d_i=k_i+1$ & arbitrary
         & $2N-3$ & topology-dependent$^{*}$ & \cite{Kiktenko2020}\\ \hline
         ${\sf iSWAP}^\bullet$ 
         & $d=3$ & linear
         & $2N-3$ & ${\cal O}(N)$ & \cite{Zheng2012}\\
         ${\sf iSWAP}^\bullet$ & $d=3$ & star
         & $\phantom{^**}2N-1^{**}$ & ${\cal O}(N)$ & \cite{Fang2023}{$^{***}$}\\
         ${\sf iSWAP}^\bullet$ & $d=3$ & arbitrary
         & $2N-3$ & topology-dependent$^{*}$ & \begin{minipage}[t]{3.2cm}\cite{Nikolaeva2022, Chu2023}\end{minipage}\\ \hline
         ${\sf XX}^\bullet(\frac{\pi}{2})$ / ${\sf XX}^\bullet(\frac{\pi}{4})$
         & $d=3$ & linear & $2N-3$ / {$4N-6$} & ${\cal O}(N)$ & \cite{nikolaeva2023universal} \\
    \end{tabular}
    
    $^*$ Ranges from ${\cal O}(N)$ for linear chains to ${\cal O}(\log N)$ for log-depth trees.
    
    $^{**}$ A clean ancillary qubit is used.

    {$^{***}$ A collective motional degree of freedom of an ion string is used as an ancilla.}
    \caption{Main characteristics of the ${\sf C}^{N-1}{\sf X}$ (${\sf C}^{N-1}{\sf Z}$) gate decompositions considered in the main text.
    }
    \label{tab:comparison}
\end{table*}

\subsection{Generalization to a multicontrolled-unitary gate}
\label{sec:gen_to_CNU}

\begin{figure}
    \centering
\includegraphics[width=\linewidth]{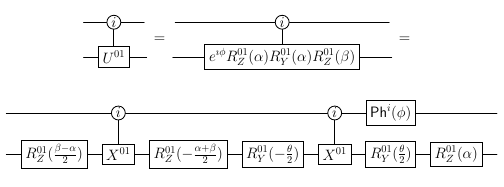}
    \caption{Construction of the central gate for realizing ${\sf C}^{N-1}U$ operation.
    The fact that arbitrary $2\times 2$ unitary $U$ can be written in the form $U=e^{\imath\phi}R_{Z}(\alpha)R_{Y}(\theta)R_{Z}(\beta)$ for some real $\phi, \alpha,\theta,\beta$ is used.
    Note that the decomposition is also valid for relative phase ${\sf CX}^{i|01}_\star$ gates, given that the input state of the target belongs to the qubit subspace.}
    \label{fig:making_CU}
\end{figure}

To conclude this section, we show how all the aforementioned decompositions of ${\sf C}^{N-1}{\sf X}$ $({\sf C}^{N-1}{\sf Z})$ gates can be generalized to a multicontrolled-unitary gate of the form
\begin{multline} \label{eq:CNU}
	{\sf C}^{N-1}U: \ket{x_1,\ldots,x_{N-1}}\ket{\psi}  \\
	\mapsto \ket{x_1,\ldots,x_{N-1}}U^{x_1\ldots x_{N-1}}\ket{\psi},
\end{multline}
where $U$ is an arbitrary $2\times 2$ unitary, $x_i\in\{0,1\}$, and $\ket{\psi}$ is an arbitrary single-qubit state.
The idea behind the generalization is in replacing the central (relative phase) ${\sf CX}^{i|01}$ or ${\sf CZ}$ gates by ${\sf C}U^{i|01}$ or ${\sf C}U$, 
respectively (here ${\sf C}U^{i|01}$ is a controlled operation that applies $U$ to the qubit subspace of the target, if the control is in $\ket{i}$).
In the latter case, ${\sf C}U$ can be realized using a well-known composition that involves no more than two ${\sf CZ}$ gates; see Corollary 5.3 in~\cite{Barenco1995}.
The generalization for relative phase ${\sf C}^{i|01}$ gate is shown in Fig.~\ref{fig:making_CU}.
Note that in the special case where $U$ has the eigenvalues $\pm1$, ${\sf C}^{N-1}U$ can be made out of ${\sf C}^{N-1}{\sf X}$ simply by surrounding the target with local basis-change operations.
We also note that a similar technique can be extended for making multicontrolled multiqubit operations [gates of the form~Eq.~\eqref{eq:CNU}, where $\ket{\psi}$ and $U$ are multiqubit state and gate, respectively].
For more technical details, see ~\cite{Kiktenko2020}.

\section{Embedding several qubits in a qudit}\label{sec:embedding_several_qubits_in_qudit}

\begin{figure}
    \centering
    \includegraphics[width=0.65\linewidth]{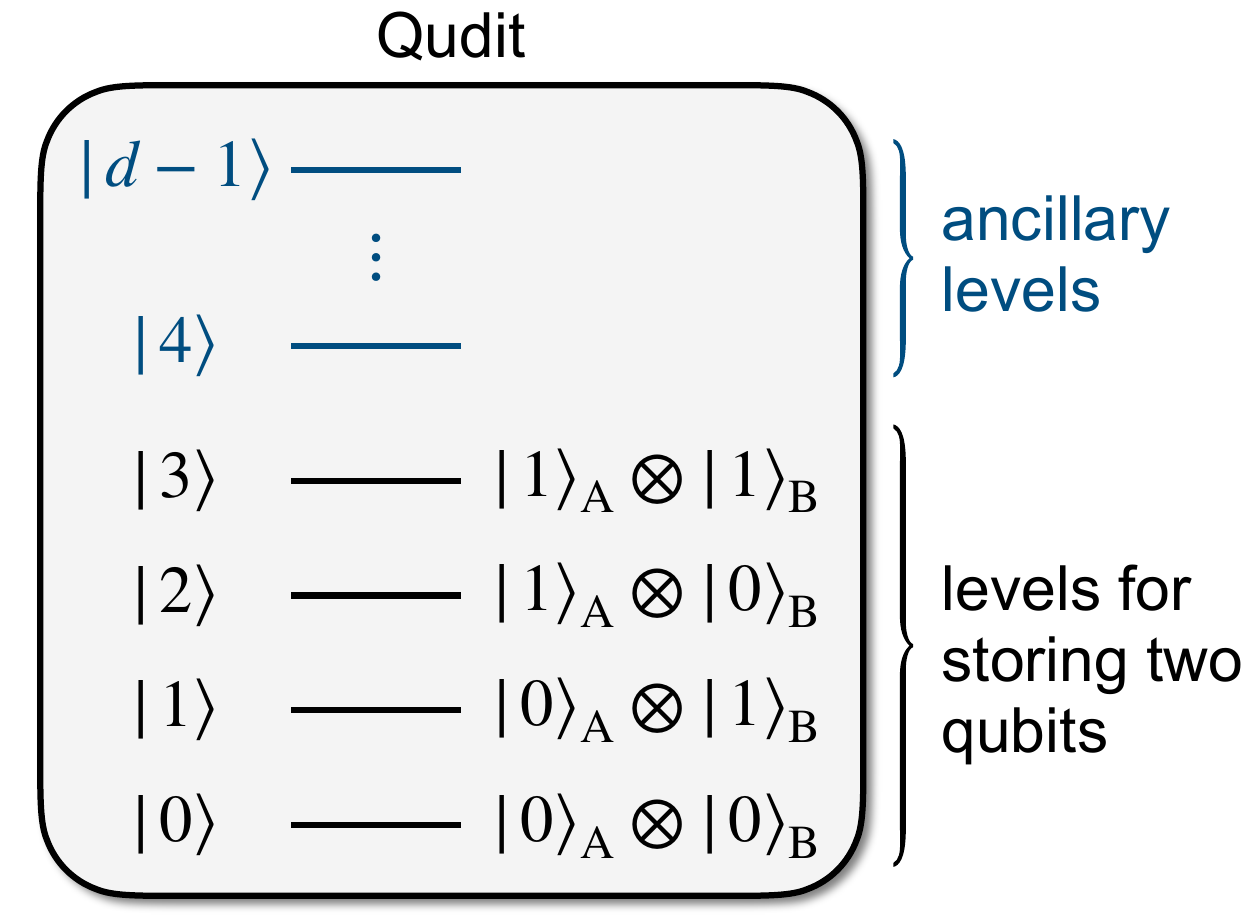}
    \caption{Example of embedding the space of $\nqb=2$ qubits A and B in the state space of a single $d$-dimensional qudit with $d\geq 5$.}
    \label{fig:qudit_as_two_qubits}
\end{figure}

Here we focus on embedding or compressing joint spaces of several qubits in spaces of distinct qudits, so we do not limit ourselves to embedding only a single qubit in each single qudit.
If the entire space of $\nqb$ qubits is placed in the state space $\mathbb{C}^d$ of a $d$-dimensional qudit, we arrive at the following decomposition:
\begin{equation} \label{eq:C_for_several_qubits}
    \mathbb{C}^d = \left(\mathbb{C}^2\right)^{\otimes \nqb} \oplus \mathbb{C}^{n_{\rm anc}},
\end{equation}
where $n_{\rm anc}=d-2^{\nqb}$ is the number of remaining ancillary levels.
A straightforward way to embed the $\nqb$ qubits' subspace in the qudit's space is to take the lowest $2^{\nqb}$ levels of the qudit and use a mapping of the form
\begin{equation} \label{eq:simple_mapping}
    \ket{k} \leftrightarrow \ket{{\rm bin}_{\nqb}(k)}, \quad k=0,1,\ldots,2^\nqb-1 ,
\end{equation}
where ${\rm bin}_{b}(x)$ is a $b$-bit string of representing $x=0,1,\ldots,2^b-1$ in the binary form [for example, ${\rm bin}_{3}(1)=0,0,1$].
See Eq.~\eqref{eq:simple_mapping} for a direct qubit-to-qudit mapping.
An example with $\nqb=2$ is shown in Fig.~\ref{fig:qudit_as_two_qubits}.

According to Eq.~\eqref{eq:C_for_several_qubits}, the dimensionality $d$ of a qudit required for storing $\nqb$ qubits grows exponentially with $\nqb$: $d \geq 2^{\nqb}$.
Taking into account apparent experimental challenges that appear when control is operater over each new level {and several theoretical issues discussed in Sec.~\ref{sec:tradeoff}}, we expect a limitation on the scalability of this approach up to the values $\nqb=2$ and 3.
Nevertheless, even using $\nqb=2$ can significantly reduce the number of two-particle operations required to execute a given circuit.
Since entangling gates commonly are the main source of decoherence, the qudit-based implementation can be advantageous, especially in the NISQ era.

We start with a discussion of the simplest example of running a two-qubit algorithm in the space of a single ququart ($d=4$) in Sec.~\ref{sec:single_qudit}, 
consider issues of coupling qubits from distinct qudits in Sec.~\ref{sec:coupling_qubits_distinct_qudits},
consider a realization of multiqubit gates for qubits in multiple qudits in Sec.~\ref{sec:multiqubit_gates_multiple_qudits}, 
discuss scaling behavior with an increase in the number of qubits embedded in a single qudit in Sec.~\ref{sec:tradeoff},
and finally discuss the opportunities and challenges coming with more general qubit-to-qudit mappings than the direct one~in~Eq.~\eqref{eq:simple_mapping} in Sec.~\ref{sec:selecting_mapping}.

\subsection{Running a qubit circuit in the space of a single qudit} \label{sec:single_qudit}

We consider a single four-level particle whose state space $\mathbb{C}^4$ is used for encoding a state of qubits A and B according to the mapping in Eq.~\eqref{eq:simple_mapping} 
(the case shown in Fig.~\ref{fig:qudit_as_two_qubits} with no ancillary levels).
Recall that, according to the model introduced in Sec.~\ref{sec:qudit-model}, 
we assume the possibility of making an arbitrary rotation $R^{ij}_\phi(\theta)$ [see Eq.~\eqref{eq:Rij}] within two-dimensional subspaces spanned by $\ket{i}$ and $\ket{j}$ 
and also the possibility of applying a single-qudit phase gate ${\sf Ph}^i(\theta)$ at any level $\ket{i}$[see Eq.~\eqref{eq:Ph_i}].
To implement a circuit designed for qubits A and B, we consider realization of single- and two-qubit gates by means of available single-qudit operations.

From the viewpoint of four-dimensional space, a single-qubit unitary
\begin{equation}
    u =
    \begin{pmatrix}
        u_1 & u_2 \\
        u_3 & u_4 \\
    \end{pmatrix},
\end{equation}
acting on A has to be considered as
\begin{equation} \label{eq:u_on_A}
    u
    \otimes
    \mathbb{1} =
    \begin{pmatrix}
        u_{1} & & u_{2} & \\
        & u_{1} & & u_{2} \\
        u_{3} & & u_{4} & \\
        & u_{3} & & u_{4} \\
    \end{pmatrix}.
\end{equation}
Similarly, if $u$ is applied to B, the resulting operation takes the form 
\begin{equation} \label{eq:u_on_B}
    \mathbb{1} \otimes u =
    \begin{pmatrix}
        u_{1} & u_{2} & & \\
        u_{3} & u_{4} & & \\
        & & & u_{1} & u_{2} \\
        & & & u_{3} & u_{4} \\
    \end{pmatrix}.
\end{equation}
According to Eqs.~\eqref{eq:u_on_A} and~\eqref{eq:u_on_B}, applying a single ``local'' gate $u$ to A (B) can be realized through a sequence of two operations: $u^{02}u^{13}$ ($u^{01}u^{23}$), 
where $u^{ij}$ denotes a $4\times 4$ unitary acting as $u$ in two-dimensional subspace spanned by $\ket{i}$ and $\ket{j}$ and acting as an identity in the remaining space.
We note that the doubling of operations appears from the necessity to avoid disturbance of the neighboring qubit in the qudit.

In contrast to a slight complication of single-qubit gate implementation, two-qubit gates can be realized by single-qudit operation.
For example, two-qubit ${\sf CZ}$ and iSWAP gates are correspondingly realized by
\begin{equation}
    {\sf Ph}^3(\pi) = \begin{pmatrix}
        1\\&1\\&&1\\&&&-1
    \end{pmatrix}, \quad
    R_0^{12}(\pi) = \begin{pmatrix}
        1 \\
        && -\imath \\
        & -\imath\\
        &&&&1
    \end{pmatrix}.
\end{equation}

According to the employed mapping in Eq.~\eqref{eq:simple_mapping}, 
the initialization in the state $\ket{0,0}$ of A and B corresponds to preparing the ququart in $\ket{0}$, while the readout measurements of A and B correspond to the ququart's measurement in the computational basis.
We note that if only A is measured, then the corresponding positive operator-valued measure can be taken as a set of two rank-2 projectors: 
$M^{\rm A}_0 = \ket{0}\bra{0}+\ket{1}\bra{1}$ and
$M^{\rm A}_1 = \ket{2}\bra{2}+\ket{3}\bra{3}$.

\begin{figure}
\centering 
\includegraphics[width=\linewidth]{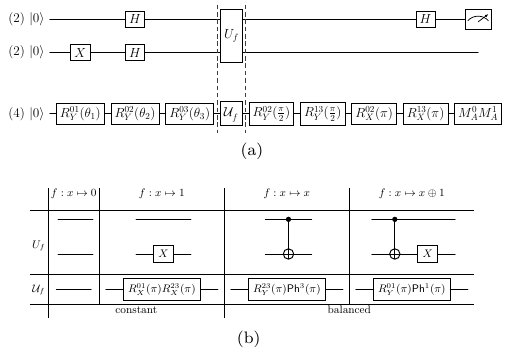}
    \caption{Implementation of Deutsch's algorithm using a single four-level system.
    (a) Correspondence between two-qubit and single-ququart circuits, where $\theta_1 = -\pi/3$, 
    $\theta_2 = 2{\arcsin}\left([2\cos(\theta_1/2)]^{-1}\right)$,
    $\theta_3 = -2{\arcsin}\left({[2\cos(\theta_1/2)\cos(\theta_2/2)]^{-1}}\right)$.
    (b) Realization of quantum oracles (in both two-qubit and single-qudit versions) for four possible bit-to-bit functions.
    Adapted from ~\cite{Kiktenko20152}}
    \label{fig:Deutsch}
\end{figure}

As a demonstration of the discussed approach, we show a realization of two-qubit circuit of Deutsch's algorithm in the space of a single ququart in Fig.~\ref{fig:Deutsch}.
Recall that the goal of the algorithm is to determine whether a bit-to-bit function $f:\{0,1\}\rightarrow \{0,1\}$ is constant or balanced given a single query to the corresponding two-qubit quantum oracle $U_f: \ket{x,y}\mapsto \ket{x,f(x)\oplus y}$. 
The answer to the problem can be read out by single-qudit $(M^{\rm A}_0, M^{\rm A}_1)$ measurement (0 represents a constant function and 1 stands for a balanced one) or by the computational basis measurement (0, 1 represents a constant function and 2, 3 stands for a balanced one).

\subsection{Coupling qubits from two distinct qudits} \label{sec:coupling_qubits_distinct_qudits}

The approach of embedding qubits in a single qudit indeed has limited scalability since the required qudit's dimensionality $d$ grows exponentially with the number of qubits.
That is why it is important to develop techniques for operating with an arbitrary number of qubits distributed in pairs, triplets, etc. over a number of qudits.
From the universal quantum computing perspective, the goal is to construct a universal qubit gate set, which can be assembled from native single- and two-qudit gates.

In Sec.~\ref{sec:single_qudit}, we have already seen how to implement arbitrary single- and two-qubit gates, which couple qubits embedded in the same qudit.
Here we show how to make coupling operations for qubits located in distinct qudits.
The crucial factor is the form of the available two-qudit entangling operation.
In what follows we focus on two cases studied in the literature: the case of generalized controlled-phase operation ${\sf CPh}^{i|j}$ (or, equivalently, the ${\sf CX}^{i|jk}$ inversion operation) and the case of the MS operation  ${\sf XX}^{ij|kl}(\chi)$.
Here we restrict our consideration to the most experimentally relevant cases of $\nqb=2$ and 3. 

\begin{figure}
\centering
\includegraphics[width=\linewidth]{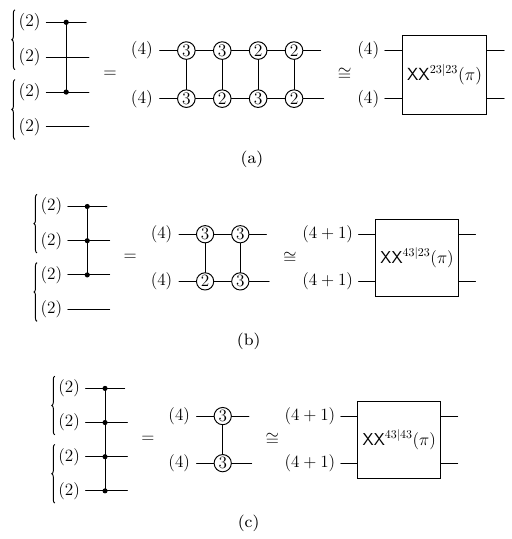}
\caption{(a) ${\sf CZ}$, (b) ${\sf C}^2{\sf Z}$, and (c) ${\sf C}^3{\sf Z}$ gates, which act on qubits from different qudits by means of ${\sf CPh}^{i|j}$ and ${\sf XX}^{ij|kl}(\pi)$ two-qudit gates.
Vertical brackets emphasize that the corresponding qubits are embedded in the same qudits.
}
\label{fig:coupling_qubits_from_qudits}
\end{figure}

Consider a system of two ququarts that contain four qubits: (A, B) in the first ququart and (C, D) in the second one (here $\nqb=2$ for both qudits).
For both ququarts we apply the direct qubit-to-qudit mapping in Eq.~\ref{eq:simple_mapping}; see Fig.~\ref{fig:qudit_as_two_qubits}.
We consider an implementation of the ${\sf CZ}$ gate acting between qubits A and C.
From the viewpoint of the four-qubit system, this corresponds to adding a $-1$ phase to four states:
$\ket{1,x,1,y}$ with all possible combinations $(x, y)\in\{0,1\}^2$.
According to the employed embedding, $\ket{1,0}\leftrightarrow\ket{2}$ and $\ket{1,1}\leftrightarrow\ket{3}$, so this operation is equivalent to four controlled-phase operations applied to four possible combinations of levels $\ket{2}$ and $\ket{3}$: ${\sf CPh}^{2|2}{\sf CPh}^{2|3}{\sf CPh}^{3|2}{\sf CPh}^{3|3}$; see also Fig.~\ref{fig:coupling_qubits_from_qudits}(a).
Using ${\sf CZ}$ at other pairs (A, D), (B, C), and (B, D) can be realized in the same way by changing the affected levels (or, equivalently, adding SWAPs acting on  qubits inside the same qudit).
The entangling ${\sf CZ}$ gate, which acts on two qubits from different ququarts, costs four entangling ${\sf CPh}^{i|j}$ operations.
We note that the origin of this overhead is similar to the one we that we discussed for qudit-based implementation of single-qubit gates.

It may be surprising that increasing the number of qubits that are affected by a gate can simplify its implementation in the qudit-based encoding.
Namely, the three-qubit gate ${\sf C}^2{\sf Z}$ acting on qubits A, B, and C, which acquires the $-1$ phase to the four-qubit states  $\ket{1,1,1,y}$ with $y\in\{0,1\}$, corresponds to two controlled-phase operations:
${\sf CPh}^{3|2}{\sf CPh}^{3|3}$ [see also Fig.~\ref{fig:coupling_qubits_from_qudits}(b)].
We note that ${\sf C}^2{\sf Z}$ (or, equivalently, the Toffoli gate) can also be used for assembling a universal gate set~\cite{Shi2003}.
Similarly, the implementation of ${\sf C}^3{\sf Z}$ on all four qubits A, B, C, and D can be realized by means of a single ${\sf CPh}^{3|3}$ gate, as shown in Fig.~\ref{fig:coupling_qubits_from_qudits}(c).
We summarize the results for ${\sf CPh}^{i|j}$-based architectures in the second column of Table~\ref{tab:two_qudits}.

\begin{table}[]
    \centering
    \begin{tabular}{c|c|c}
         Qubit gate & No. of ${\sf CPh}^{\bullet}$s & 
         No. of ${\sf XX}^\bullet(\pi)$ {/ ${\sf XX}^\bullet(\frac{\pi}{4})$} \\\hline
         ${\sf CZ}$ (same qudit)& 0 & $0~/~0$\\
         ${\sf CZ}$ (distinct qudits)& 4 & $1~/~{4}$\\
         ${\sf C}^2{\sf Z}$ & 2 & $1~/~{4}$ \\
         ${\sf C}^3{\sf Z}$ & 1 & $1~/~{4}$ \\
    \end{tabular}
    \caption{The number of qudit entangling gates required to realize a given entangling gate applied to (a subset of) four qubits embedded in two qudits according to the direct qubit-to-qudit mapping in Eq.~\eqref{eq:simple_mapping}.
    {The presented numbers of ${\sf XX}^\bullet$ gates required for the decomposition of ${\sf C}^2{\sf Z}$ and ${\sf C}^3{\sf Z}$ operations are applicable for qudit systems with a dimension $d\geq 5$.}
    }
    \label{tab:two_qudits}
\end{table}

We then consider the construction of multiqubit gates based on an ${\sf XX}^{ij|kl}(\chi)$ entangling operation.
To begin, we note that ${\sf XX}^{23|23}(\pi)$ acquires the $-1$ phase to four states 
\begin{equation} \label{eq:XXpi}
    \begin{aligned}
        &\ket{2,2}\leftrightarrow \ket{1,0,1,0}, \quad 
        \ket{2,3}\leftrightarrow \ket{1,0,1,1},\\     
        &\ket{3,2}\leftrightarrow \ket{1,1,1,0}, \quad 
        \ket{3,3}\leftrightarrow \ket{1,1,1,1}
    \end{aligned}
\end{equation}
(the qudits' states are on the left and the corresponding qubits' ones are on the right).
This is exactly ${\sf CZ}$ between the qubits A and C; see Fig.~\ref{fig:coupling_qubits_from_qudits}(a).
Having ${\sf CZ}$ act on other qubit pairs (A-D, B-C, and B-D) is realizable by adding interqudit SWAP gates.
In contrast to ${\sf CPh}^{i|j}$, there is no overhead in the number of entangling gates: a single ${\sf CZ}$ entangling operation between two qubits from distinct qudits costs a single entangling operation between these qudits.

We then consider an implementation of ${\sf C}^{2}{\sf Z}$ and ${\sf C}^{3}{\sf Z}$ qubit gates with an ${\sf XX}^{ij|kl}(\chi)$ operation.
On the one hand, it is always possible to employ some known qubit-based decomposition of ${\sf C}^{2}{\sf Z}$ or ${\sf C}^{3}{\sf Z}$ down to single- and two-qubit gates, such as those of ~\cite{Barenco1995, saeedi2013linear, nakanishi2021quantumgate}, and then to implement it on qudits by replacing two-qubit ${\sf CZ}$ gates with either local or two-qudit gates.
Note that the resulting number of entangling gates becomes less than for a qubit-based realization due to the transformation of some qubit ${\sf CZ}$ gates into single-qudit operations.
On the other hand, as shown by~\cite{nikolaeva2023universal}, ${\sf C}^{2}{\sf Z}$ and ${\sf C}^{3}{\sf Z}$ can be realized with a single ${\sf XX}^{ij|kl}(\pi)$ operation given one ancillary unpopulated level.
In particular, ${\sf C}^{2}{\sf Z}$ on A, B, C, and ${\sf C}^{3}{\sf Z}$ on A, B, C, and D are realized in the qubits' subspaces (spanned by $\ket{x,y}$, $x,y\in\{0,\ldots,3\}$) by applying
${\sf XX}^{43|23}(\pi)$ and ${\sf XX}^{43|43}(\pi)$ operations, respectively; see also Fig.~\ref{fig:coupling_qubits_from_qudits}(c, b).
The idea behind this implementation is to place a necessary number of $-1$ elements in qubit's subspace (two $-1$ elements for ${\sf C}^{2}{\sf Z}$ and one $-1$ element for ${\sf C}^{3}{\sf Z}$) by moving other $-1$ elements to the remaining subspace. [Recall that according to Eq.~\eqref{eq:XXpi}, ${\sf XX}^{ij|kl}(\pi)$ is a diagonal matrix with four $-1$ elements.]
In this way a single ancillary level allows one to compose all possible ${\sf C}^n{\sf Z}$ gates ($n=1,2,3)$, acting on four qubits embedded in pairs in two qudits, with no more than a single ${\sf XX}^{ij|kl}(\pi)$ operation; see the right column in Table~\ref{tab:two_qudits}.

\begin{figure}
	\centering
\includegraphics[width=\linewidth]{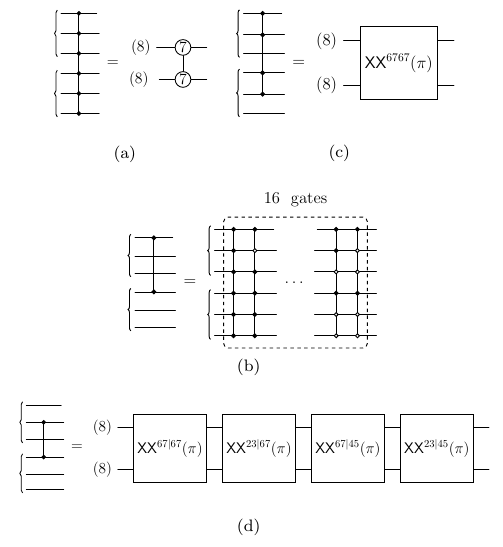}
    \caption{Realization of gates acting on qubits embedded in triples in qudits.
    (a) Equivalence between ${\sf C}^5{\sf Z}$ and ${\sf CZ}^{7|7}$.
    (b) Scheme for making a two-qubit ${\sf CZ}$ with 16 ${\sf CPh}^{i|j}$ gates. 
    The standard convention of denoting inverse control with empty dots is followed.
    (c) Realization of ${\sf C}^3{\sf Z}$ via ${\sf XX}^{ij|kl}(\pi)$  gates.
    (d) Scheme for making a two-qubit ${\sf CZ}$ gate using ${\sf XX}^{ij|kl}(\pi)$ gates.}
    \label{fiq:quocts}
\end{figure}

To conclude this subsection, we consider the case of using $d=8$ qudits (quocts) for embedding triples of qubits ($\nqb=3$) according to Eq.~\eqref{eq:simple_mapping}.
To begin, we note that a single ${\sf CPh}^{7|7}$ gate, applied to two qudits, realizes ${\sf C}^5{\sf Z}$ to all six qubits in these qudits; see Fig.~\ref{fiq:quocts}(a).
This comes from the fact that $\ket{7}\leftrightarrow\ket{1,1,1}$.
At the same time, the realization of a ${\sf CZ}$ gate for two qubits from distinct qudits requires $2^4=16$ ${\sf CPh}^{i|j}$ operations; see Fig.~\ref{fiq:quocts}(b).
This is the price for not disturbing four ``neighboring'' qubits embedded in these qudits.
A single ${\sf XX}^{67|67}(\pi)$ operation corresponds to a ${\sf C}^3{\sf Z}$ gate acting on four of six qubits, as shown in Fig.~\ref{fiq:quocts}(c).
This leads to the implementation of ${\sf CZ}$ gate for qubits from different qudits with four ${\sf XX}^{ij|kl}(\pi)$ operations; see Fig.~\ref{fiq:quocts}(d).

In conclusion, we see that in the case of  ${\sf CPh}^{i|j}$- (${\sf CX}^{i|jk}$-)based platforms, increasing the number of qubits $\nqb$ embedded in a single qudit affects the resulting number of entangling gates in a nontrivial way.
On the one hand, it allows some of the two-qubit gates from the original qubit circuit to be replaced with local qudit gates. On the other hand, there is an increasing overhead in the number of entangling gates for coupling qubits from different qudits.
In Sec.~\ref{sec:tradeoff} we return to a discussion of the strengths and weaknesses of embedding multiple qubits within a single qudit.

\subsection{Applying multiqubit gates to qubits in multiple qudits} \label{sec:multiqubit_gates_multiple_qudits}

In Sec.~\ref{sec:coupling_qubits_distinct_qudits}, we obtained a universal gate set, which consists of single- and two-qudit gates and allows one to realize a qubit circuit.
Here, by employing techniques from Sec.~\ref{sec:qudit-assisted_decs}, we show how upper ancillary levels $i=2^{\nqb},\ldots,d-1$, if available, can be used for simplifying decompositions of gates acting on a number of qubits distributed over more than two qudits.
To demonstrate a general idea, we restrict ourselves to the case of ${\sf C}^{2N-1}{\sf Z}$ gates acting on $2N$ qubits embedded (or compressed) in pairs in $N$ qudits of dimension $d\geq 5$ according to the qubit-to-qudit mapping in Eq.~\eqref{eq:simple_mapping}.
We assume that no qubits other than those affected by the gate are embedded in these $N$ qudits.
See~\cite{Nikolaeva2021} for a discussion of more general cases.

From the viewpoint of qubit states, the ${\sf C}^{2N-1}{\sf Z}$ gate realizes an operation of the form
\begin{equation} \label{eq:2NCZ}
    \ket{x_1,\ldots,x_{2N}} \mapsto (-1)^{x_1\ldots{x_{2N}}} \ket{x_1,\ldots,x_{2N}}.
\end{equation}
Assuming that $(2k)$th and $(2k+1)$th qubits are embedded in the same qudit ($k=1,\ldots,N$), and taking into account that  $\ket{11}\leftrightarrow\ket{3}$, we come to the following 
transformation in the qudits' space:
\begin{equation} \label{eq:y1yn}
    \ket{y_1,\ldots,y_{N}} \mapsto (-1)^{\delta_{y_1,3}\dots \delta_{y_1,3}} \ket{y_1,\ldots,y_{N}},
\end{equation}
where $\delta_{a,b}$ stands for the Kronecker symbol.
Notably, there is a clear similarity between Eq.~\eqref{eq:y1yn} and the operation of the ${\sf C}^{N-1}{\sf Z}$ gate. The only difference is that the phase factor is acquired by the all-3s state $\ket{3}^{\otimes N}$ rather than $\ket{1}^{\otimes N}$.
This is the key observation that allows us to derive the ${\sf C}^{2N-1}{\sf Z}$ decomposition using the results in Sec.~\ref{sec:qudit-assisted_decs}.

\begin{figure*}
    \centering
\includegraphics[width=\linewidth]{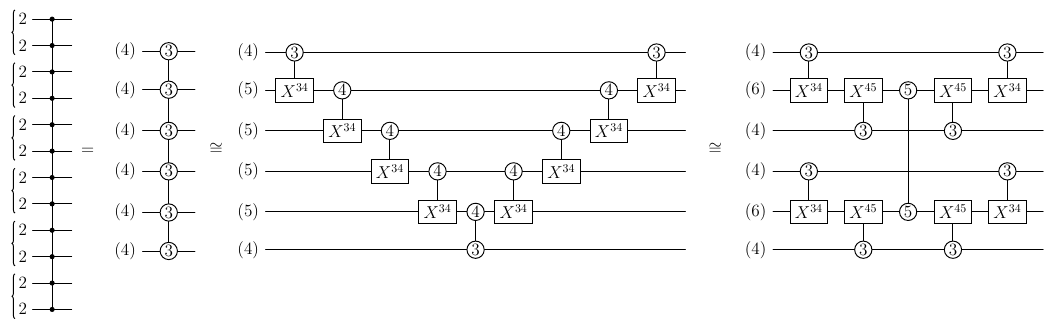}
\caption{
Two decompositions of a ${\sf C}^{11}{\sf Z}$ gate acting on $12$ qubits embedded in pairs in six qudits. A trade-off between the resulting circuit depth and the number of leveraged extra levels is demonstrated.
}
    \label{fig:C2NZququarts}
\end{figure*}

In Fig.~\ref{fig:C2NZququarts} we show two decompositions based on ${\sf CX}^{i|jk}$ gates.
The first V-shaped decomposition requires qudits of dimension $d\geq 5$ coupled within a linear-chain topology.
The second decomposition requires qudits of dimension $d\geq 6$ (two ancillary levels are required) and is realized within a binary-tree coupling map.
Both decompositions require $2N-3$ of entangling gates, yet the second one benefits from logarithmic depth.

The decomposition of Fig.~\ref{fig:C2NZququarts} can be adapted to ${\sf XX}^{ij|kl}(\chi)$-based platforms.
The key idea is to employ the trick shown in Fig.~\ref{fig:coupling_qubits_from_qudits}(c) to transform ${\sf XX}^{ia|ia}(\pi)$ into ${\sf CPh}^{i|i}$ using an ancillary level $\ket{a}$ (recall that ${\sf CPh}^{i|i}$ can be transformed into ${\sf CX}^{i|jk}$ by means of local gates).
The ancillary level $\ket{a}$ cannot be used for other purposes.
Thus, the V-shaped decomposition from Fig.~\ref{fig:C2NZququarts} can be realized with ${\sf XX}^{ij|kl}(\pi)$ gates and qudits of dimension $d\geq 6$ (one ancillary level is used for transforming ${\sf XX}$ into ${\sf CPh}$, while the other is used for making the decomposition itself), and the binary-tree one can be realized using qudits of dimension $d\geq 7$.
The properties of resulting decompositions are summed up in Table~\ref{tab:C2NZ}.
Ancillary levels serve as a useful resource for making quantum computations:
each new level under control provides an advantage in multiqubit gate decomposition.
We note that there is a way of implementing the considered ${\sf C}^{2N-1}{\sf Z}$ gate using $(24N-18)$ ${\sf XX}^{ij|kl}(\pi)$ operations acting on ququints ($d=5$) coupled within a linear chain.
The idea is to realize a phase-relative version of ${\sf CPh}^{i|j}$ with ${\sf XX}^{ij|kl}(\pi)$ without assisting any ancillary levels by adopting a known decomposition of a qubit ${\sf C}^4{\sf Z}$ gate.
A single ancillary level V-shaped scheme is then employed. 
For more details see ~\cite{nikolaeva2023universal}.

\begin{table*}[]
    \centering
    \begin{tabular}{c|c|c|c|c|c}
        Entangling gate & Qudit dimension $d$ & Coupling map & No. of entangling gates & Circuit depth & Reference \\ \hline\hline
        ${\sf CX}^{i|jk}$ & 5 & Linear & $N-3$ & ${\cal O}(N)$ & \cite{Nikolaeva2023ququints}\\
        ${\sf CX}^{i|jk}$ & 6 & Binary tree & $N-3$ & ${\cal O}(\log N)$ & \cite{Kiktenko2020} \\ \hline
        ${\sf XX}^{\bullet}(\pi)$ {~/~${\sf XX}^{\bullet}(\frac{\pi}{4})$} & 6 & Linear & $N-3$ {~/~$4N-12$} & ${\cal O}(N)$ & \multirow{2}{4cm}{ \cite{nikolaeva2023universal}} \\
        ${\sf XX}^{\bullet}(\pi)$ {~/~${\sf XX}^{\bullet}(\frac{\pi}{4})$} & 7 & Binary tree & $N-3$~/~ {$4N-12$} & ${\cal O}(\log N)$ & \\
    \end{tabular}
    \caption{Comparison of properties of decompositions of a ${\sf C}^{2N-1}{\sf Z}$ gate that acts on $2N$ qubits embedded in pairs over $N$ qudits.}
    \label{tab:C2NZ}
\end{table*}

{
\subsection{Scaling behavior with an increase in the number of qubits embedded in a single qudit} \label{sec:tradeoff}

Here we address some points related to the increase in the number of qubits $\nqb$ embedded within a single qudit.
We focus on two illustrative examples, the so-called best- and worst-case scenarios, from the viewpoint of a qudit-based realization of qubit circuits. 
For a more in-depth technical discussion, see~\cite{Nikolaeva2021}.
For simplicity, we fix the ${\sf CPh}^\bullet$ operation as the basic qudit entangling gate, which can be realized via ${\sf XX}^{\bullet}(\pi)$ with the aid of an extra ancillary qudit level according to the approach shown in Fig.~\ref{fig:coupling_qubits_from_qudits}(c).

As discussed, increasing the number of qubits in qudit $\nqb$ allows for a reduction in the total number of physical particles required to implement an $n$-qubit circuit to {\color{violet}$\lceil n/\nqb \rceil$}.
At the same time, the minimum qudit dimension required grows exponentially ($d\geq 2^{\nqb}$), which represents a significant experimental challenge.
However, there are several additional considerations regarding the implementation of qubit gates in qudits, which we now discuss in more detail.

We begin by considering an implementation of single-qubit gates.
When previously discussed cases of $\nqb=2, 3$ are generalized, applying a single-qubit unitary $u$ turns into applying a $u\otimes \mathbb{1}^{\otimes (\nqb-1)}$ gate in the entire qubits subspace for the corresponding qudit (here the position of $u$ in the tensor product is determined by the index of the affected qubit among all $\nqb$ qubits embedded in the qudit). 
After expanding the Kronecker product, one obtains a unitary matrix consisting of $2^{\nqb-1}$ two-level operations, as is shown for $\nqb=2$ in Eqs.~\eqref{eq:u_on_A} and~\eqref{eq:u_on_B}.
Thus, the overhead in the number of two-level single-qudit operations required to implement a single-qubit gate grows exponentially with $\nqb$.
We emphasize that, from an experimental point of view, there is no absolute need to be able to address all possible $2^{\nqb-1}(2^{\nqb}-1)$ pairs of transitions in order to perform the required operation.
Any connected graph of transitions, for example, a ladderlike one, is sufficient, even though it necessitates the introduction of extra two-level swapping pulses.

We now turn to the implementation of a ${\sf C}^{N-1}{\sf Z}$ gate, including the two-qubit $N=2$ case.
From the viewpoint of several-qubits-in-one-qudit schemes, the cost of implementing ${\sf C}^{N-1}{\sf Z}$ largely depends on how $N$ affected qubits are distributed among the qudits.
The main conceptual principle is that the lower the number of involved distinct qudits, the lower the cost in terms of the number of entangling gates.
The most favorable situation for the qudit-based implementation is when all $N\leq \nqb$ qubits are embedded in the same qudit.
In this case performing the ${\sf C}^{N-1}{\sf Z}$ operation is equivalent to applying a sequence of $2^{\nqb-N}$ single-qudit ${\sf Ph}^i(\pi)$ phase operations~[Eq.~\eqref{eq:Ph_i}].
The levels $i$ in the sequence are chosen in such a way that the qubit states encoded in the qudit's state $\ket{i}$ satisfy a ``mask'' like $\ket{1,1,*,1}$, with $N$ 1s and $(\nqb-N)$ asterisks: 1s are located on qubits affected by the ${\sf C}^{N-1}{\sf Z}$, and an asterisk represents all possible bit values on other positions.
In the given example, two phase operations with $i=13\leftrightarrow \ket{1,1,0,1}$ and $i=15\leftrightarrow \ket{1,1,1,1}$ are required.
Given the ability to perform phase operations in a virtual manner, i.e., without applying actual physical pulses to the system, it is reasonable to anticipate that the implementation of $2^{\nqb-2}$ phase gates will be faster and of better quality than the physical implementation of a two-qubit and multiqubit gate within the straightforward qubit-based realization.

In the case of two qudits involved in ${\sf C}^{N-1}{\sf Z}$ operation (with $2\leq N\leq 2\nqb$), qudit entangling gates become necessary.
In particular, one has to apply a number of ${\sf CPh}^{i|j}$ operations, with $\ket{i}\otimes\ket{j}$ satisfying a specific two-qudit binary mask, for example, $\ket{*,1,*,*}\otimes\ket{*,1,*,1}$.
As before, $N$ 1s indicate positions of qubits that are affected by the ${\sf C}^{N-1}{\sf Z}$ and $2\nqb-N$ asterisk symbols represent arbitrary possible values of remaining qubits.
Here a favorable situation for qudit-based implementation is  ${\sf C}^{2\nqb-1}{\sf Z}$ acting on \emph{all $2\nqb$ qubits} in two qudits, which requires a single entangling ${\sf CPh}^{i|j}$ gate to the $i=j=2^{\nqb-1}$ levels.
At the same time, an unfavorable two-qubit ${\sf CZ}$ gate on qubits from distinct qudits requires $2^{2\nqb-2}$ entangling ${\sf CPh}^{i|j}$ gates.

We also consider two additional scenarios: a favorable one, where a ${\sf C}^{K\nqb-1}{\sf Z}$ gate operates on all $K\nqb$ qubits embedded in $K$ qudits and at least one extra level in each qudit is available, and the most unfavorable situation, where ${\sf C}^{N-1}{\sf Z}$ acts on ${N}$ qubits from $N$ different qudits in such a way that each qudit contains $\nqb-1$  unaffected qubits with no auxiliary levels available in qudits. 
In the first scenario, by extending the scheme shown in Fig.~\ref{fig:C2NZququarts}, we obtain a decomposition consisting of $2K-3$ ${\sf CPh}^{i|j}$ gates.
In the second scenario,  we have no other option but to employ a standard qubit-based approach and implement each ${\sf CZ}$ gate using $2^{2\nqb-2}$ ${\sf CPh}^{\bullet}$ operations, thereby obtaining a $2^{2\nqb-2}$-fold increase in the number of entangling gates.

In Fig.~\ref{fig:tradeoff} we present the number of entangling gates for all considered cases (excluding those that did not require any entangling gates at all). The number of ${\sf CZ}$ (${\sf CX}$) gates in a qubit-based implementation is taken from~\cite{Maslov2016}.
In particular, the cheapest of the number of ${\sf CX}$ gates clean-ancilla-based decompositions are considered.
Figure~\ref{fig:tradeoff} captures how the number of entangling gates required for the implementation of ${\sf C}^{N-1}{\sf Z}$  varies significantly depending on the distribution of $N$ qubits among qudits. 
We see that qudit-based implementations may offer both substantial advantages and disadvantages compared to conventional qubit-based solutions.
In this way we expect that increasing the number of qubits ($\nqb$) can be promising for running a certain class of quantum algorithms, with extensive use of multiqubit gates.
A notable example is the qudit-based Grover algorithm, as discussed by~\cite{Nikolaeva2023ququints}.
At the same time, increasing $\nqb$ may lead to challenges in terms of the number of entanglement gates required for running unstructured quantum circuits or circuits that initially consist only of single-qubit and two-qubit gates, such as those used in experiments aiming to demonstrate quantum advantage~\cite{Martinis2019, Morvan2023, Pan2021-4,Pan2021-5}. 
Note also that the measurement of the population of qudit levels corresponds to a simultaneous (computational basis) measurement of all $\nqb$ qubits embedded within this qudit, which should be taken into account during the transpilation process.

\begin{figure}
    \centering
    \includegraphics[width=\linewidth]{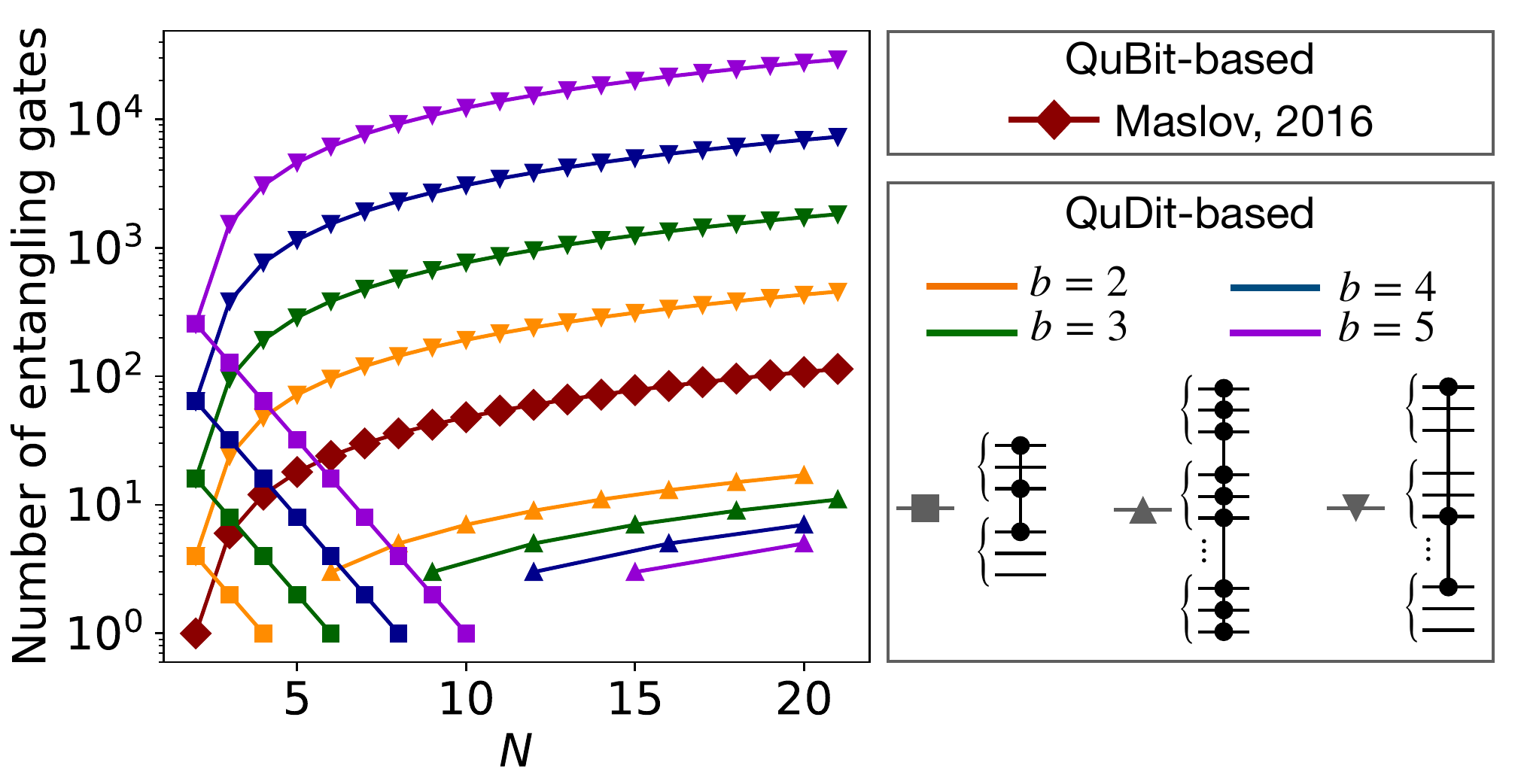}
    \caption{Number of entangling gates, ${\sf CPh}^\bullet$ for qudit-based architectures 
    and ${\sf CZ}$ or ${\sf CX}$ for the qubit-based one, required for making a ${\sf C}^{N-1}{\sf Z}$ gate.
    In the qudit-based case, the results for different values of the number of qubits $\nqb$ embedded in one qudit are shown.
    The results for the scenario, where $2\leq N\leq 2\nqb$ qubits are embedded into two distinct qubits, are denoted by squares $\square$.
    Up triangles $\triangle$ represent the case where $N = K\nqb$ qubits are distributed among $K$ qudits, while down triangles $\triangledown$ represent the application of ${\sf C}^{{N-1} {\sf Z}}$ to $N$ qubits from $N$ distinct qudits.}
    \label{fig:tradeoff}
\end{figure}
}

\subsection{Varying a qubit-to-qudit mapping}
\label{sec:selecting_mapping}

\begin{figure}
    \centering
\includegraphics[width=\linewidth]{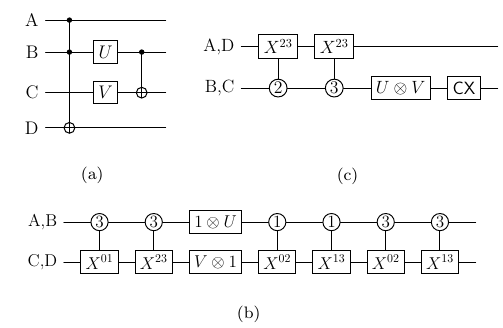}
    \caption{Example of the situation where two different embeddings of four qubits in two ququarts ($d=4$) results in qudit circuits with different depth and numbers of two-qudit ${\sf CX}^{i|jk}$ gates. (a) Original qubit circuit. (b), (c) Qudit circuits corresponding to different embeddings.}
    \label{fig:non-equivalent_mappings}
\end{figure}

Here we discuss some points related to the question of how qubits can be embedded in qudits.
To begin we recall that all decompositions from Secs.~\ref{sec:coupling_qubits_distinct_qudits} and \ref{sec:multiqubit_gates_multiple_qudits}
are designed with respect to the particular qubit-to-qudit mapping in Eq.~\eqref{eq:simple_mapping}.
It fixes the correspondence between computational basis states of pairs or triples of qubits and levels of a qudit; however, there is still ambiguity as to which qubits of the original circuit should be placed in which qudits.
In Fig.~\ref{fig:non-equivalent_mappings} we show an illustration of the fact that different pairings of four qubits over two ququarts can yield vastly different resulting qudit circuits, especially in the case of ${\sf CX}^{i|jk}$- (${\sf CPh}^{i|j}$-) based platforms.
Namely, for the four-qubit circuit shown in Fig.~\ref{fig:non-equivalent_mappings}(a), changing a mapping can result in a reduction of the number of two-particle gates from six~[Fig.~\ref{fig:non-equivalent_mappings}(b)] to two~[Fig.~\ref{fig:non-equivalent_mappings}(c)].
Although both qudit circuits realize the same qubit circuit, one can expect a higher fidelity for the circuit in Fig.~\ref{fig:non-equivalent_mappings}(b) than for the one in Fig.~\ref{fig:non-equivalent_mappings}(c).

Finding the best possible mapping (for example, from the viewpoint of the number of entangling gates or the resulting circuit depth) generally seems to be a difficult computational problem. 
In the case where $m$ all-to-all coupled qudits embed exactly $n=\nqb m$ qubits (i.e., each of the $m$ qudits is used for storing $\nqb$ qubits), the number of nonequivalent mappings is given by
\begin{equation}
    N_{\rm maps} = \frac{(\nqb m)!}{m! (\nqb!)^m},
\end{equation}
which grows superexponentially with $m$ for $\nqb>1$ (note that for $\nqb=1$, $N_{\rm maps}=1$).
We call two mappings equivalent if they differ by permutation of qubits inside a qudit or permutation of two qudits with the same contained qubits.
These permutations do not affect the resulting number of two-qudit gates within the all-to-all coupling map or circuit depth.

From a practical perspective, it is not always essential to find the {optimal} mapping in order to benefit from the use of qudit-based qubit circuits.
A suboptimal mapping obtained via a heuristic algorithm can already provide an advantage over a straightforward qubit-based realization of a given qubit circuit.
In particular, if the number of available qudits $m$ is not less than the number of qubits $n$ of the processed qubit circuit, then there is an option to employ the qudit-based processor as a qubit-based one
(at least within our model from Sec.~\ref{sec:qudit-model}) .
At the same time, if the processed qubit circuit contains some multicontrolled unitaries, its implementation with the assistance of upper levels 
(using techniques from Sec.~\ref{sec:qudit-assisted_decs}) definitely results in a decrease of the number of two-particle gates and is expected to improve the resulting fidelity.
This consideration raises the problem of developing an efficient classical software tool kit for transpiling (compiling) qubit circuits for existing qudit-based hardware platforms.

The first steps in this direction were presented by ~\cite{Nikolaeva2021, Mato2023graphs}.
\cite{Nikolaeva2021} considered an approach that tried all possible nonequivalent mappings, transpiling a given input qubit circuit according to each mapping and choosing the one with the smallest number of ${\sf CPh}^{i|j}$ entangling gates.
Note that transpilation is performed independently of each qubit-to-qudit mapping.
\cite{Mato2023graphs} proposed another method in which all multiqubit gates are first decomposed down to a single-qubit and two-qubit gates and the task of distributing qubits among qudits is then considered.
The key idea in this method is to use a graph representation of a circuit where the graph's nodes are given by the qubit levels, edges are given by the local and nonlocal operations, and weights of the edges correspond to a number of corresponding gates.
Then the task of association nodes with the highest weights of connections, using an adaptation of the $k$-means algorithm, is considered.
In the result a decomposition with ${\sf CPh}^{i|j}$ entangling gates is obtained.
This approach removes the necessity of an exhaustive search among possible mappings yet does not benefit from all the advantages brought by qudits for decomposing multiqubit gates.

\begin{figure*}
\centering
\includegraphics[width=\linewidth]{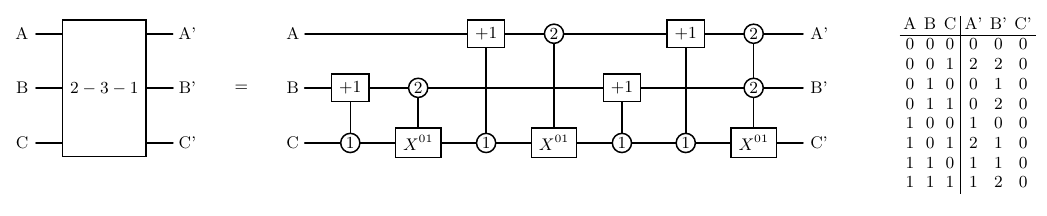}
    \caption{Scheme for obtaining a clean ancilla on C' by compressing three qubits into two qutrits. 
    A compression circuit and a truth table are shown. 
    The employed controlled +1 gates denote an operation $\ket{i}\mapsto \ket{i+1~(\text{mod}~3)}$ on the target given a control in the state $\ket{1}$.
    Adapted from~\cite{Baker2020}.
    }
    \label{fig:compression}
\end{figure*}

The second point we discuss here relates to the fact there is no restriction requiring  the space of a qudit to be used only for embedding the entire space of one or several qubits.
\cite{Baker2020} considered a compression scheme for using pairs of qutrits of the total dimension $3^2=9$ for encoding triples of qubits of the dimension $2^3=8$.
The circuit for such compression and  the corresponding truth table are given in Fig.~\ref{fig:compression}.
This scheme allows a clean ancilla on the third quantum information carrier (C') to be obtained.
\cite{Baker2020} proposed using these ancillas within the realization of an adder.
We note that although ancillas represent a valuable resource for realizing quantum computing, applying logical operations to compressed qubits becomes a nontrivial problem, that was not addressed in the original  proposal.

\section{Experimental progress in the development of qudit-based platforms}
\label{sec:exp_progress}

Most physical platforms considered for use in quantum computing are capable of operating in both qubit and qudit modes.
Here we provide an overview of recent experimental results regarding the development of large-scale qudit processors, whereas seminal works on basic principles of operating with qudits within corresponding physical systems were acknowledged in the Introduction. 

{\it Neutral atoms and ions.}
Atoms (both neutral and charged) are natural candidates for qudit-based computing owing to the presence of a number of discrete energy levels that can be used for storing quantum information.
Ions also benefit from the strong Coulomb interaction that makes it possible to engage a motional degree of freedom to couple internal states.
Recently, two groups have demonstrated experimental control trapped-ion-based qudit systems; see ~\cite{Ringbauer2021,Kolachevsky2022}.
The processor presented by~\cite{Ringbauer2021} uses $^{40}{\rm Ca}^+$ ions and supports up to 10 ions with $d=7$ levels in each. 
The qudit states are encoded in Zeeman sublevels of $4^{2}S_{1/2}$ and $3^{2}D_{5/2}$ levels, and each $S$-$D$ transition is accessed via a single narrow-band laser.
Randomized benchmarking (RB) of the qutrit and ququint Clifford operations shows error rates per Clifford operation of
$2(2)\times10^{-3}$ and $1.0(2)\times10^{-2}$, respectively.
As basic entangling gates, controlled-exchange and controlled-increment operations based on the MS gate were considered. 
Corresponding estimated fidelities of $97.5(2)\%$ and $93.8(2)\%$ were reported.
The two-ququart processor demonstrated by Ref.~\cite{Kolachevsky2022} was based on operations with $d=4$ levels in $^{171}{\rm Yb}^{+}$.
For storing quantum information, it employs Zeeman sublevels of the quadrupole clock transition $6{^2}S_{1/2} \leftrightarrow 5{^2}D_{3/2}$.
Single-ququart fidelities obtained by population estimation vary from 83\% to 89\%, depending on the levels pair, whereas the two-qudit MS has been realized with an estimated fidelity of $65(4)\%$.
Recently, an upgraded version of the system was used to implement ${\sf C}^2{\sf Z}$, ${\sf C}^3{\sf Z}$, and ${\sf C}^4{\sf Z}$ gates with truth table fidelities of $95.7(2)\%$, $88.0(2)\%$, and $80.0(2)\%$,  respectively~\cite{nikolaeva2024ions} .
In these experiments, the scheme from Fig.~\ref{fig:Toffoli_linear}(b) has been adapted for dual-type optical-microwave qutrits, allowing only global control over the ancillary levels.

Noteworthy results in increasing the number of controlled levels were demonstrated by~\cite{Anderson2015}, where unitary operations on $d=16$ qudit associated with the $6S_{1/2}$ ground state of $^{133}{\rm Cs}$ achieving fidelities larger than $98\%$ are shown.
\cite{Low2023} recently reported an average state preparation and measurement (SPAM) fidelity of $91.7(3)\%$ for operations with $d=13$ levels of $^{137}{\rm Ba}^+$.

Realizations of ${\sf C}^2{\sf X}$, ${\sf C}^3{\sf X}$, and ${\sf C}^4{\sf X}$ gates in $^{171}{\rm Yb}^+$-based system using a qutrit version of a Cirac-Zoller gate with corresponding truth table fidelities of $95.4(4)\%$, $92.2(5)\%$, and $90.4(6)\%$ have been demonstrated~\cite{Fang2023}.
Notably, in the presented decomposition, the motional degree of freedom served as an ancilla within the discussed linear-depth scheme shown in Fig.~\ref{fig:Toffol_star}(b).

In addition to the MS and Cirac-Zoller gates, another type of two-qubit entangling operation that is native to the trapped-ion platform is the light-shift (LS) gate. 
The protocol with this gate has recently been generalized for qudits of various dimensions that are encoded in the Zeeman sublevels of $^{40}$Ca$^{+}$ ions~\cite{Hrmo2023}. 
The main advantage of the LS-gate-based scheme  is that it can generate entanglement between multiple qudit levels in a single application, whereas other protocols typically require multiple applications to achieve the same goal. Moreover, an experimental realization of a qudit LS-based entangling gate with high fidelity was demonstrated: gate fidelities of $99.6(1)\%, 98.7(2)\%, 97.0(2)\%,$ and $93.7(3)\%$ were obtained for $d = 2, 3, 4,$ and 5, respectively~\cite{Hrmo2023}.
The use of the qudit LS gate for the realization of qubit circuits with qudits is a promising avenue for further research.

Neutral atoms and ion-based qudits are also successfully used for quantum simulations.
~\cite{Meth2023} presented experimental simulations of two-dimensional lattice quantum electrodynamics with matter fields using a system of single-qutrit ($d=3$) plus four qubits ($d=2$) and a system of three ququints ($d=5$)  within $^{40}{\rm Ca}^+$-based platform.
\cite{Kazmina2024demonstration} used five qutrits $(d=3)$ in an array of ten ions to demonstrate a parity-time symmetry breaking. 
We also note a proposal for a hardware-efficient quantum simulator of non-Abelian gauge theories based on Rydberg atom qudits~\cite{Zache2022} that was suggested for use in simulating the dynamics of general gauge theories coupled to matter fields~\cite{Zache2023, Zoller2023}.

{\it Superconducting platform.}
The basic idea behind quantum computing with superconducting circuits is in processing quantum states of anharmonic oscillators realized by Josephson junctions.
The lowest two levels of oscillators are used to make qubits; however, upper levels can also be utilized.
{Significant progress has been made in the use of transmons as qutrits ($d=3$), while there is also an increasing trend toward the use of higher levels~\cite{Ustinov2015}.
Operating with $d=4$~\cite{Lidar2024, Goss2023Toffoli, Kehrer2024},~$d=8$ \cite{Champion2024}, and $d=12$ levels~\cite{wang2024d12transmon} in transmons have been demonstrated.}
The main challenge in this direction is related to decreasing the coherence times for upper levels compared to the lower ones, which limits the involved dimensionality $d$.

{Owing to the remarkable theoretical and experimental progress in the implementation of RB techniques for qudit gates~\cite{Sanders2020rb, Sanders2024qtrb, Sanders2024rb, Sanders2024rb2, Siddiqi2021, Goss2023Toffoli, Goss2023RC}, RB became a widespread tool for characterizing single-qudit gates on superconducting processors. 
First performed on qutrits \cite{Siddiqi2021} and then on ququarts \cite{Goss2023Toffoli}, it shows that the fidelity of single-qudit gates is approaching that of single-qubit gate operations.
For instance, ~\cite{Goss2023Toffoli} found that the average native gate fidelities for qudits with dimensions $d=2$, $d=3$, and $d=4$ were found to be $99.936(3) \%$, $99.909(4) \%$ and $99.78(2) \%$, respectively.
\cite{Champion2024} demonstrated that when treating a $(d=8)$-level transmon as a 7/2-spin system, the displacement operator and a virtual selective number-dependent arbitrary phase (SNAP) gate can be used to implement  arbitrary single-qudit unitary operations with $\mathcal{O}(d)$ physical pulses.
The average Clifford gate fidelities were shown to be $99.51(2)\%$, $98.86(2)\%$, and $98.25(6)\%$ for $d=3$, $5$, and $8$, respectively, while the displacement pulse fidelities were estimated to be $99.71(1)\%$, $99.31(1)\%$, and $98.95(3)\%$ for the same values of $d$.

A significant contribution to the reduction in the fidelity of the qudit Clifford gate is made by the increase in the number of two-level rotations required to implement Clifford gates as the value of $d$ increases.
In addition, there is an increase in the size of the Clifford group with respect to $d$.
For example, the single-qutrit Clifford group consists of 216 elements modulo a global phase~\cite{Siddiqi2021}; for comparasion, a single-qubit Clifford group contains 24 elements.
For this reason, in certain experiments a ``qubitlike-'' RB is employed to estimate the fidelity of single-qudit gates, where operations are characterized in only two-dimensional subspaces~\cite{wang2024d12transmon}. 
For $d=12$, the process infidelities of qubit-like gates were reported to be $3\cdot 10^{-3}$~\cite{wang2024d12transmon}. Benchmarking was conducted in each adjacent level qubit subspace at the lowest ten levels. 
For this system the fidelity of the state $2^{-1/2} (\ket{0} + \ket{8})$ was reported to be $98.2\%$.
The main factors that limit the fidelity of states and operations in such a highly complex system are shorter coherence times at higher levels, a greater number of errors in physical pulses, and SPAM errors.}

Significant progress in increasing the fidelity of entangling two-qutrit gates, namely, $\widetilde{\sf CZ}: \ket{j,k} \mapsto (e^{\imath 2\pi/3})^{j,k}\ket{j,k}$, was shown in by~\cite{Goss2022}: 
estimated process fidelities of $97.3(1)\%$ and $95.2(3)\%$ were achieved for $\widetilde{\sf CZ}^\dagger$ and $\widetilde{\sf CZ}$, respectively.
{Process fidelity of a two-qutrit gate has been obtained with the use of cycle benchmarking~\cite{Siddiqi2021}.
A fidelity of $89.3(1.9)\%$ was reported for a two-qutrit controlled-phase gate as a result of the quantum process tomography experiment described by~\cite{Yu2023}. 
}

{Pioneering experiments aimed at improving qudit gate performance and error mitigation techniques have also been conducted on a superconducting platform. 
These include dynamical decoupling experiments for $d=3$ and $d=4$ qudits ~\cite{Iiyama2024, Lidar2024} and randomized compiling, which, in combination with noise-free output extrapolation, allowed a reduction in the infidelity of a three-qudit Greenberger–Horne–Zeilinger (GHZ) state by a factor of 3~\cite{Goss2023RC}. Following this, the state fidelity of the purified density matrix was reported to be $99.8 \%$
We note that the first successful preparation of a three-qudit GHZ state using a superconducting platform was demonstrated in~\cite{Galda2022} with a fidelity of $76(1)\%$.
This example illustrates the impact of error suppression techniques and improvements in qudit operation fidelity over time in experimental research.}

The decomposition of the Toffoli gate on the IBM quantum superconducting quantum processor \texttt{ibmq\_jakarta} with the help of the third level and the use four entangling gates within a linear-chain topology was reported by~\cite{Galda2021}.
\cite{Wallraff2012} showed an increase in gate fidelity in comparison with the pioneering work on a qutrit-assisted Toffoli gate decomposition on superconducting platform~: $68.5(5)\% \rightarrow 78.00 \pm 1.93\%$.
A further improvement in three-qubit gate fidelity was demonstrated by~\cite{Hill2021}, who realized a double-controlled-phase gate with the fidelity of $87.1(8)\%$ on three particles of a 32-qutrit Aspen-9 processor.
Notably, an ${\sf iSWAP}^{01|12}(\theta)$ operation has been used as a basic entangling gate.
{The ${\sf iSWAP}^{ij|kl}$-based scheme shown in Fig.~\ref{fig:folding-unfolding} has been experimentally implemented on a superconducting processor to demonstrate the $N$-qubit Toffoli gates with $N = 4, 6$, and 8, as reported by \cite{Chu2023}.
The reported truth table basis state fidelity values are $89.1\%, 53.2\%,$ and $39.1\%$.
Based on this decomposition, the Grover search algorithm has also been implemented for 64 items.
Furthermore, the fidelities of $96.0(3)\%$ and $92(1)\%$ for ${\sf C}^2{\sf Z}$ and ${\sf C}^3{\sf Z}$ gates were demonstrated via cycle benchmarking by~\cite{Goss2023Toffoli}, who utilized a two-qutrit cross-Kerr gate-based decomposition.
}

A two-qutrit transmon-based processor was also announced by~\cite{Roy2023}.
In this processor local single-qutrit gates are realized by applying pulses at $\ket{0}\leftrightarrow\ket{1}$ and $\ket{1}\leftrightarrow\ket{2}$ transitions, and an entanglement is created via a generalized controlled-phase gate.
The processor presented by~\cite{Roy2023} has been used for implementing ternary versions of Deutsch-Jozsa, Bernstein-Vazirani, and Grover’s search algorithms.

{In addition to algorithms, superconducting qudits are useful for simulation purposes.
For example, the Fermi-Hubbard model was recently simulated using $d=4$ ququarts \cite{Vezvaee2024fh}.
}
Furthermore, a single superconducting qutrit has also been used in a demonstration of the PT-symmetry breaking (together with the aforementioned trapped-ion system)~\cite{Kazmina2024demonstration}.

{\it Photonics platform.} 
Photonic architectures are also prominent for qudit-based computing owing to the variety of degrees of freedom inherent to photons; see ~\cite{Zeilinger2018} for a review.
Notably, the first qudit-assisted decomposition of multiqubit gates was shown with quantum light~\cite{White2009}; a Toffoli gate with a truth table overlap of $81(3)\%$ was shown.
We also note that the first two-qutrit GHZ state preparation with a fidelity of $75.2( 2.88)\%$ was obtained with an encoding state in the orbital-angular momentum of photons~\cite{Erhard2018}.
In a more recent experiment~\cite{Bao2023photons}, a four-photon three-dimensional GHZ state $\ket{\rm GHZ}_4^3=3^{-1/2}(\ket{0000}+\ket{1111}+\ket{2222})$ was generated with a fidelity of $72.2(1.8)\%$, which is greater than the lower bound of 2/3, thus confirming the entanglement of the state.
In a programmable two-qudit integrated-optics system with a dimension of the qudits of up to $d=15$, considered by~\cite{Wang2018photons}, the maximally entangled state fidelity obtained using quantum state tomography has been found to be $96\%$, $87\%$, and $81\%$ for $d = 4, 8$, and 12, respectively.

Recently, the first photonic two-ququart ($d=4$) processor~\cite{Wang2022} was reported; it uses path encoding within an integrated circuit.
Classical statistical mean fidelities of $98.8(1.3)\%$ and $96.7(1.9)\%$ have been reported for Pauli and Fourier single-ququart gates.
One of the features of this processor is the use of a multiqudit, multivalue controlled-unitary gate that applies a $d$-ary gate to the target qudit depending on the state of the control qudit. 
Quantum process fidelity has been reported at $95.2\%$ for a two-ququart gate of this type.
Moreover, a single time-bin-encoded photonic qubit of dimension $d = 32$ was utilized to factorize the number 15 using Shor's algorithm, as described by~\cite{Weng2024}.
A successful preparation of entangled photonic angular momentum states with quantum numbers up to 10010 was reported by~\cite{Fickler2016}.

{\it Promising platforms for qudit-based computing.}

Other promising platforms for qubit-based quantum computing include high-spin systems such as molecules and molecular magnets \cite{Hernándezantón2024, Luis2020Gd2, Chiesa2024}, electron-nuclear states in a donor in silicon \cite{Morello2024nature, Morello2024-1}, and nitrogen-vacancy (NV) centers \cite{Lukin2024,Du2024nv}.
These systems provide a large number of levels [up to 8 for the GdW$_{30}$ molecule~\cite{Hernándezantón2024}, 16 for the $^{123}$Sb donor in silicon~\cite{Morello2024nature}, and 64 for the Gd$_2$ molecule~\cite{Luis2020Gd2}]. Typically, their Hilbert space can be treated as a space with several qubits (three qubits for the GdW$_{30}$, four qubits for the $^{123}$Sb donor, and six qubits for GD$_2$). 
Recent experiments have demonstrated that it is possible to efficiently control such single-qudit systems. 
For instance, the average single-qubit gate fidelity of the $^{123}$Sb donor qubit, as estimated using gate set tomography on a selected pair of qubit levels out of a total of 16 available, was higher than 99.4\%~\cite{Morello2024nature}.

Notably, the first theoretical proposal to implement Grover's search was described for a single molecular magnet considered as a qudit by \cite{Loss2001}. 
Afterward, it was experimentally implemented using a single nuclear 3/2 spin carried by a Tb ion, sitting in a single molecular magnet transistor \cite{Balestro2017}. 
A qutrit NV-center-based spin system was used to verify quantum contextuality~\cite{Du2024nv}.
There are several proposals on how to embed a logical qubit in a single spin qudit \cite{Chiesa2024, Gross2021, Morello2024-1}. 

Ultracold molecules can also be used for qudit-based quantum computing since they have very rich level structure. In some cases, they can contain up to 96 levels~\cite{Sawant2020}.
The recent demonstration of entangling two-particle operations between CaF molecules~\cite{Bao2023molecules, Holland2023} is a significant step forward in the development of qudit processors based on this technology. 
Notably, as a basic entangling operation, a two-particle iSWAP gate was implemented between two qubits encoded in molecular rotational states.

\section{Discussion and conclusion}
\label{sec:concl}

\begin{figure*}
    \centering
    \includegraphics[width=0.9\linewidth]{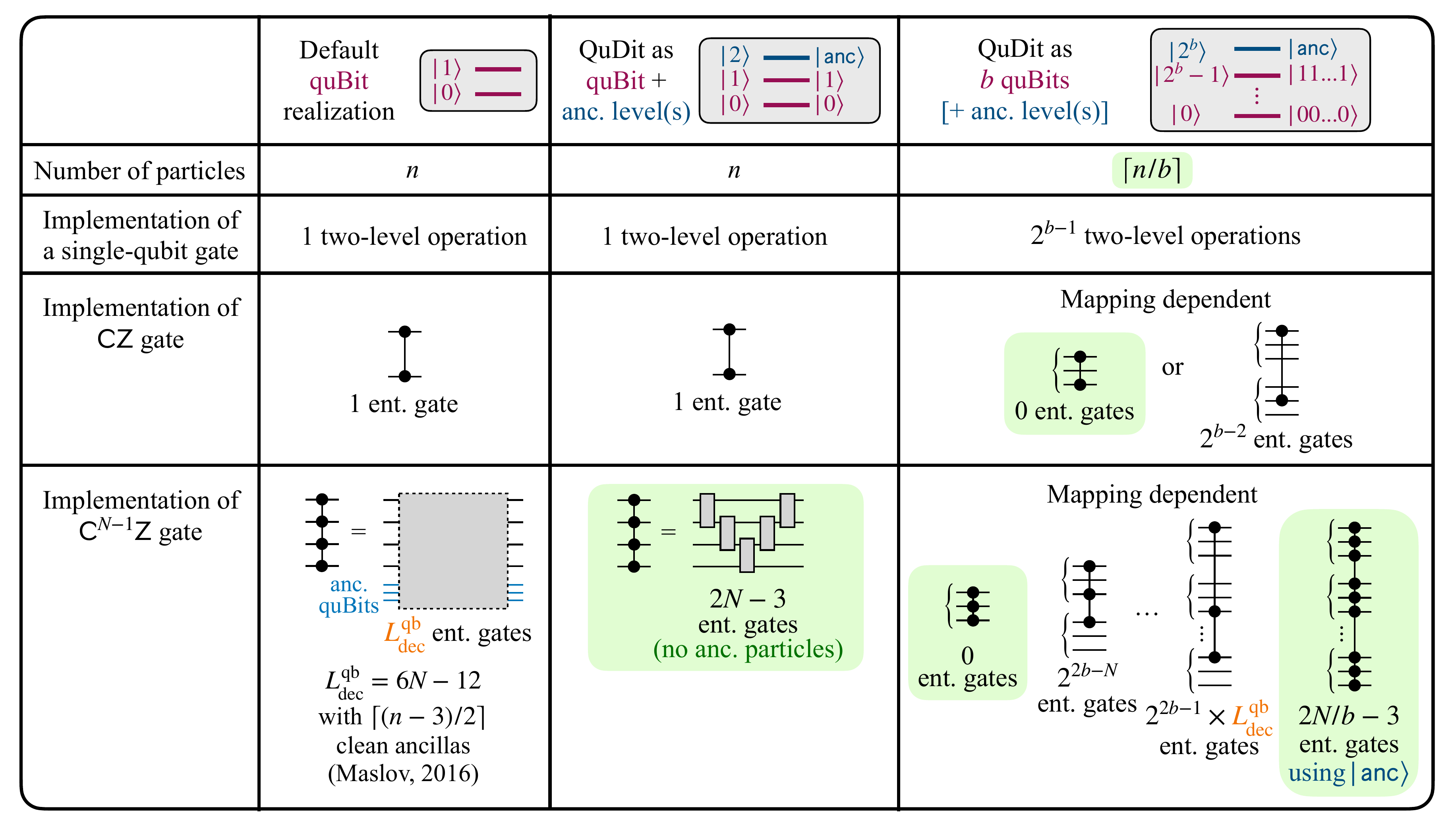}
    \caption{{Schematic comparison of the main approaches for running qubit-based circuits considered in the text.
    The situations where qudit-based implementation provides considerable advantage over the straightforward qubit-based implementation are highlighted.
    For simplicity, only the results of ${\sf CPh}^{\bullet}$ qudit entangling  gates are presented.}
    }
    \label{fig:table}
\end{figure*}

Here we summarize the reviewed approaches to using qudit-based hardware for running quantum circuits that were initially represented in qubit form.
{The main features of all the approaches and their comparison with the straightforward qubit-based realization are presented in Fig.~\ref{fig:table}.}
First, we have explored the possibility of simplifying the implementation of multiqubit gates by utilizing higher levels of qudits as a temporary information buffer, {while maintaining the implementation of two-qubit gates using a conventional qubit-based approach.}
Although the general concept of all the considered decompositions is similar, 
we have shown that the concrete realization may vary significantly depending on the available entangling operations, the connectivity topology, and the accessible dimensionality of the qudits.
{For all the considered entangling operations, it is sufficient to have only one additional level, that is, $d = 3$, in order to achieve a decomposition of an $N$-qubit Toffoli gate with only $2N-3$ entangling gates [${\sf CPh}^{\bullet}$, ${\sf iSWAP}^\bullet(\pm \frac{\pi}{2})$, or ${\sf XX}^\bullet(\frac{\pi}{2})$].
For comparasion, the best known qubit-based decompositions require $6N + {\rm const}$ entangling gates~\cite{Maslov2016} with clean ancillas or $4N + {\rm const}$ entangling gates with measurement-based feedforward operations~\cite{He2017}).}

Second, we have considered the possibility of using qudits of dimension $d\geq 4$ for embedding (compressing) $\nqb=\lfloor \log_2 d \rfloor$ qubits.
We have demonstrated that it is possible to implement a $\nqb$-qubit circuit using a single qudit by rewriting various single-qubit and two-qubit gates as single-qudit operations.
In the scenario where $n$ qubits are embedded within $\lceil n/\nqb \rceil$ qudits, it has been observed that the complexity of implementing ${\sf C}^{N-1}{\sf Z}$ multiqubit operations (including the case of $N=2$) is significantly influenced by the manner in which the affected $N$ qubits are arranged across the qudits. 
Depending on this arrangement, there may be an advantage or a disadvantage in terms of the number of entangling operations required compared to a straightforward qubit-based implementation.
The most favorable scenario for implementing  ${\sf C}^{N-1}{\sf Z}$ using qudits is when the $N$ qubits are grouped together into as few qudits as possible, such as when $N = K\nqb$ qubits are combined into $K$ qubits. 
Conversely, the least favorable case would be where the $N$ qubits are spread across $N$ separate qudits.
The techniques of simplifying multiqubit operations developed for the case where each qudit  is embedded with only a single qubit can be efficiently combined with the technique of compressing several qubits in a qudit 
[in this case, $2^{\nqb}$ levels are used to store qubits and the remaining $(d-2^{\nqb})$ as ancillary ones].
We have also discussed some techniques and issues related to tuning a qubit-to-qudit mapping.
On the one hand, we have shown that changing the mapping ``on the fly'' can result in getting a clean ancillary quantum information carrier. On the other hand, we have faced the nontrivial problem of finding the (sub)optimal mapping.

To provide a comprehensive picture, we have made a review of some recent experimental results on developing qudit-based hardware platforms.
We have seen that the fidelity of operations with qudits becomes comparable to that for standard qubit-based architectures, which provides motivation for further theoretical research in the field of qudit-based computing.
We expect that all the discussed schemes can be realized with upcoming generations of qudit processors.

Finally, we emphasize a set of open issues and research directions that are important for the further development of universal qudit-based processors and their application in qubit algorithms.
The first such issue is that a deeper understanding of the connection between the available type of two-qubit interactions and the design of two-qubit gates for qubits embedded in these qudits is required.
The open question is a construction of a universal scheme for a general two-qudit interaction.

The second significant topic is the investigation of the impact of noise within the discussed schemes.
In particular, there is an interesting challenge in developing error-correction methods~\cite{Devitt2013} for the case of multiple qubits embedded in a single qudit.
Note that local errors in a single qudit can result in entangling errors between qubits contained within that qudit.
We anticipate that the development of such error-correction techniques should take into account the specific structure of the available transitions and other physical factors, such as crosstalk, in a given physical realization of qudits~\cite{Gross2021}, as well as the specific physical mapping between qubit states and qudit states, especially when multiple physical degrees of freedom for a single particle are utilized~\cite{Jia2024}.
Certainly, the continued development of qudit syndrome decoding techniques is also important~\cite{Hutter2015}.

The third open challenge is to find the efficient methods for (i) embedding qubits within the qudit space and (ii) allocating qubits among qudits within the implementation of a given qubit circuit. 
We anticipate that more efficient methods of qubit embedding than those given by Eq.~\eqref{eq:simple_mapping} could be developed for a specific hardware platform.
An example of a comparison of different physical mapping techniques within the ion-trap-based platform was given by~\cite{shivam2024}.
An efficient algorithm will also be required for finding a generally (sub)optimal distribution of qubits among qudits, which optimizes a specified cost function, such as the number of entangling gates or the circuit depth.
The development of such an algorithm will be essential for realizing the full potential of qudit systems for executing qubit quantum circuits.

\section*{Acknowledgments}

We thank N.N. Kolachevsky, A.A. Mardanova, I.A. Semerikov, and I.V. Zalivako for their valuable comments. 
Our research is supported by the Priority 2030 program at NUST ``MISIS'' under Project No. K1-2022-027.
The work of A.S.N. and E.O.K. was also supported by RSF Grant No.~19-71-10091 (review of qutrit-based and ququart-based decompositions).
A.K.F. also acknowledges the RSF Grant No.~19-71-10092 (analysis of quantum algorithm implementation). 

\bibliography{bibliography-qudits.bib}

\end{document}